\documentclass[twocolumn,twocolappendix]{aastex631}

\usepackage{ascmac}
\usepackage{amsmath}
\usepackage{amssymb}
\usepackage{amsfonts}
\usepackage{bm}
\usepackage[T1]{fontenc}

\begin{document}

\title{A Novel Method to Constrain Tidal Quality Factor from A Non-synchronized Exoplanetary System}

\author[0000-0003-1579-5937]{Takato Tokuno}
\affiliation{ Department of Astronomy, School of Science, The University of Tokyo, 7-3-1 Hongo, Bunkyo-ku, Tokyo 113-0033, Japan }
\affiliation{ School of Arts \& Sciences, The University of Tokyo, 3-8-1 Komaba, Meguro, Tokyo 153-8902, Japan }
\email{tokuno-takato@g.ecc.u-tokyo.ac.jp}

\author[0000-0002-4909-5763]{Akihiko Fukui}
\affiliation{Komaba Institute for Science, The University of Tokyo, 3-8-1 Komaba, Meguro, Tokyo 153-8902, Japan}
\affiliation{Instituto de Astrof\'isica de Canarias, V\'ia L\'actea s/n, E-38205 La Laguna, Tenerife, Spain}

\author[0000-0001-9734-9601]{Takeru K. Suzuki}
\affiliation{ Department of Astronomy, School of Science, The University of Tokyo, 7-3-1 Hongo, Bunkyo-ku, Tokyo 113-0033, Japan }
\affiliation{ School of Arts \& Sciences, The University of Tokyo, 3-8-1 Komaba, Meguro, Tokyo 153-8902, Japan }
\affiliation{Komaba Institute for Science, The University of Tokyo, 3-8-1 Komaba, Meguro, Tokyo 153-8902, Japan}

\begin{abstract}
We propose a novel method to constrain the tidal quality factor, $Q'$, from an observed non-synchronized star-planet system consisting of a slowly rotating low-mass star and a close-in Jovian planet, taking into account the co-evolution of stellar spin and planetary orbit due to the tidal interaction and the magnetic braking. On the basis of dynamical system theory, the track of the co-evolution of angular momentum from the fast rotator regime for such a system exhibits the existence of a forbidden region in the $\Omega_\mathrm{orb}$ -- $\Omega_\mathrm{spin}$ plane , where $\Omega_\mathrm{spin}$ and $\Omega_\mathrm{orb}$ denote the angular velocity of the stellar spin and planetary orbit, respectively. The forbidden region is determined primarily by the strength of the tidal interaction. By comparing ($\Omega_\mathrm{orb},\Omega_\mathrm{spin}$) of a single star-planet system to the forbidden region, we can constrain the tidal quality factor regardless of the evolutionary history of the system. The application of this method to the star-planet system, NGTS-10 -- NGTS-10 b, gives $Q' \gtrsim 10^8$, leading to an tight upper bound on the tidal torque. Since this cannot be explained by previous theoretical predictions for non-synchronized star-planet systems, our result requires mechanisms that suppress the tidal interaction in such systems.
\end{abstract}

\keywords{Exoplanet systems (484) --- Exoplanet tides (497) --- Star-planet interactions (2177) --- Low mass stars (2050) --- Hot Jupiters (753) --- Analytical mathematics (38)}


\section{Introduction} \label{sec:Introduction}
Thanks to recent space telescope missions such as \textit{Kepler} \citep{Borucki2010Sci} and TESS \citep{Ricker2014SPIE} and to the follow-up missions using ground telescopes, a large number ($\gtrsim$ 5500) of exoplanets have been detected. A significant fraction of them is orbiting a low-mass main-sequence star that possesses a convective envelope, similarly to our solar-system planets. High-quality observational data enable us to obtain precise stellar, planetary, and orbital parameters of numerous star-planet systems \citep[e.g., NASA Exoplanet Archive;][]{Akeson2013PASP} and to reveal unexpectedly great diversity of star-planet systems \citep[see, e.g.,][]{Winn2015ARAA,Jontof-Hutter2019AREPS}. 

One of the representative examples that are not common in the Solar system is close-in Jovian planets. The firstly discovered exoplanet orbiting a solar-like star was such a ``hot Jupiter'', 51 Pegasi b \citep{MayorQueloz1995Nat}, and since then, hot Jupiters have been observed at a high frequency, whereas there is a caveat that hot Jupiters are preferentially identified by the radial velocity and planetary transit techniques. They are commonly considered to have formed beyond the ice line and migrated inward to their current positions \citep[see, e.g.,][]{DawsonJohnson2018ARAA, Fortney2021JGR}. The migration of planets is induced by interactions with the central star, disk, and other planets/companions \citep[see, e.g.,][Section 10.10 and 10.11]{Perryman2018}. 

In particular, the tidal interaction between a low-mass star and a close-in Jovian exoplanet is one of the key mechanisms of the star-planet interaction. The tidal dissipation in the stellar interior, which we focus on hereafter, plays an important role in the last phase of planetary migration. Specifically, when the stellar spin is slow, the tidal interaction transports the orbital angular momentum (AM) to the stellar spin, reducing the semi-major axis and spin-orbit inclination angle of the planetary orbit \citep[][]{Darwin1880RSPT,Zahn1977AA, Hut1981AA, Barker2009MNRAS}. While the orbital decay is confirmed for only one system, WASP-12 b \citep[e.g.,][]{Yee2020ApJL}, low-inclination trends in close-in companions orbiting around a low-mass star are acquired statistically \citep{Winn2010ApJ,Albrecht2012ApJ, Albrecht2022PASP}. Also, the AM transport causes a spin-up of the host star, and this tendency is observationally obtained by the statistical examination \citep{Tejada-Arevalo2021ApJ}. In addition to the spin-up effect, the tidal interaction has also attracted attention in terms of stellar magnetism because it may increase stellar magnetic activity \citep{Cuntz2000ApJ}. However, despite the importance, quantitatively consistent treatment of the efficiency of the tidal interaction has yet to be established because of both theoretical and observational difficulties.

First, the tidal interaction is a complex problem that is composed of multiple nonlinear processes \citep[e.g.,][see also Section \ref{subsubsec:tide_torque}]{Ogilvie2014ARAA}, making self-consistent quantitative discussion a challenging task. Additionally, the low-mass stars also lose their AM through the magnetic braking effect \citep{WeberDavis1967ApJ, Kawaler1988ApJ, Matt2012ApJ, Shoda2020ApJ}.  When considering both the magnetic braking effect and tidal interaction, the co-evolution of AM is expected to exhibit complicated behaviors \citep{Dobbs-Dixon2004ApJ, Barker2009MNRAS, Damiani2015AA, Gallet2019AA, Benbakoura2019AA}. 

Second, it is extremely difficult to directly constrain the theoretically obtained tidal strength. In order to do so, we have to know the evolutionary history of star-planet systems, at least the age of systems and the initial distribution of planets \citep[e.g.,][]{Penev2014PASP,Penev2018AJ}. While the stellar age, which roughly corresponds to the age of the system, is basically determined by isochrones of stellar model grids \citep[e.g.,][]{Baraffe2015AA,Choi2016ApJ}, \citet{Tayar2022ApJ} showed that the derived age includes systematic uncertainty of tens of percent and that, furthermore, model dependence cannot be ignored. In addition to that, recent observations by ALMA reveal diverse nature of protoplanetary disks and their evolution, suggesting that it is a herculean task to pin down initial distributions of planets in different systems \citep[see, e.g.,][for detail]{Drazkowska2023ASPC,Miotello2023ASPC}. It is difficult to remove the indecisiveness of results due to these uncertainties.

The main purpose of this paper is to investigate the method to restrict the efficiency of tidal interaction without facing the above uncertainties. Then, we focus on the non-synchronized system composed of a slowly-rotating low-mass star and a close-in Jovian planet in order to simplify the physical picture. Also, considering the differential equations governing the AM co-evolution from a mathematical viewpoint of the dynamical system, we propose a novel method to impose a constraint on the efficiency of tidal interaction. A great advantage of our method is that we require only minimal assumptions about the model of tidal interaction and about the age and initial condition of the system, respectively.

The outline of this paper is as follows. In Section \ref{sec:Method}, we first describe the formulation of AM co-evolution and propose a novel method to constrain the tidal interaction using the dynamical system concept. After the brief introduction of observational data (Section \ref{sec:Observation}), we show the main obtained by applying our method to observations (Section \ref{sec:Result}). Discussions and conclusions are presented in Sections \ref{sec:Discussion} and \ref{sec:Conclusion}, respectively.

\section{Method}
\label{sec:Method}

In this section, we describe the methodology of our approach. We first introduce our framework (Section \ref{subsec:Framework}). By giving some assumptions, we formulate to track the AM co-evolution of the star-planet system. Next, we explain how to constrain the efficiency of tidal interactions using the mathematical concept of the dynamical system (Section \ref{subsec:Concept}). Finally, we describe the specific procedures for application to observations in Section \ref{subsec:Procedure}.

Before starting the analysis, we make a remark about our target system. We set the two-body system formed by a low-mass main-sequence star and a close-in Jovian planet orbiting it. We consider the time evolution after the zero-age main sequence (ZAMS), which is defined here as the stage when the stellar spin is fastest as a result of the contraction \citep[e.g.,][]{Bouvier1997AA, Gallet2013AA}. The typical lifetime of protoplanetary disks is $\lesssim 10$ Myr \citep[][]{Ribas2015AA, Kunitomo2020MNRAS}, which is shorter than the duration of the pre-main sequence phase of low-mass stars before reaching the ZAMS. Therefore, we ignore the effect of the migration driven by disk-planet interaction. 

We also assume that both stellar and planetary properties, such as mass and radius, except for the rotation are constant with time for simplification. In fact, this assumption is not strictly correct. A low-mass main-sequence star generally expands and brightens with evolution. Detailed stellar evolution calculations  shows that the lower the stellar mass is, the smaller the variation rate of the stellar parameter is. On the other hand, while planetary mass and radius may decrease through atmospheric escape \citep[e.g.,][]{Owen2019AREPS}, it is considered that the Jovian planets can maintain their envelope \citep[e.g.,][]{Kurokawa2014ApJ}. On the basis of the above discussions, this assumption is reasonable for the system that consists of a low-mass central star and a massive gaseous planet. In this paper, we consider a K or M-type star, $M_\mathrm{s} \leq 0.8 \, M_{\odot}$, and a gas-giant planet with $M_\mathrm{p} \geq 0.5 \, M_\mathrm{J}$, where $M_\mathrm{s}$, $M_\mathrm{p}$, $M_\odot$ and $M_\mathrm{J}$ represent the stellar mass, planetary mass, Solar mass and Jovian mass, respectively. Quantitative verification of the errors caused by this treatment is presented in Section \ref{subsec:validity}. We note that, hereafter, the suffixes ``s'', ``p'', ``$\odot$'', and ``J'' represent the variables of star, planet, the Sun and the Jupiter, respectively.

\subsection{Framework for AM evolution} \label{subsec:Framework}

\subsubsection{Basic equations} \label{subsubsec:equations}

\begin{figure}[t!]
\centering
\plotone{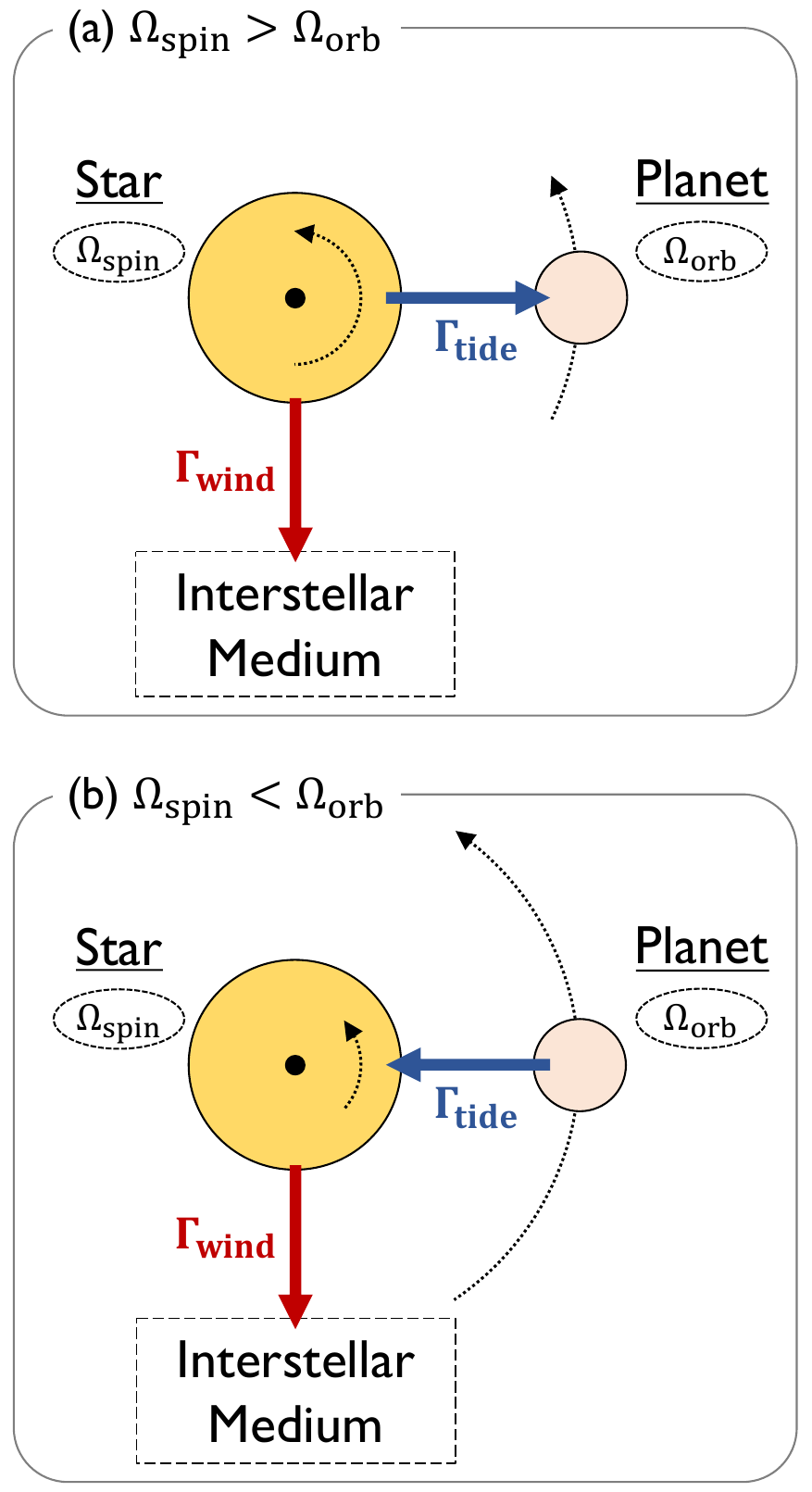}
\caption{Schematic picture of AM co-evolution in a star-planet system. The red and blue arrows represent the AM transport through the magnetic braking and the tidal interaction, respectively. In (a), the planet is orbiting outside the co-rotation radius. Since $\Omega_\mathrm{spin} > \Omega_\mathrm{orb}$ in this case, the AM of the stellar rotation is transferred to the orbital AM by the tidal interaction. In (b), the planet is inside the co-rotation radius and $\Omega_\mathrm{spin} < \Omega_\mathrm{orb}$; the orbital AM is transferred to the spin AM by the tidal interaction. In both cases, the magnetic braking always removes the AM of the system.
\label{fig:Schematic1}}
\end{figure}

The time evolution of AM is driven by the tidal interaction and magnetic braking. While the central star loses the spin AM by the magnetic braking, the tidal interaction, which is driven by an internal force, exchanges the spin AM of the star, 
\begin{align}
J = I_\mathrm{s} \Omega_\mathrm{spin}, \label{eq:AM_assume1}
\end{align}
and the orbital AM, 
\begin{align}
h = M_\mathrm{s} M_\mathrm{p} G^{2/3}(M_\mathrm{s}+M_\mathrm{p})^{-1/3} \Omega_\mathrm{orb}^{-1/3}, \label{eq:AM_assume2}
\end{align}
where $I_\mathrm{s}$, $\Omega_\mathrm{spin}$, $\Omega_\mathrm{orb}$, and $G$ denote the stellar moment of inertia, the angular velocity of stellar spin, the orbital angular velocity, and the gravitational constant, respectively. 
In equation (\ref{eq:AM_assume1}) we are assuming that the stellar spin is rigid\footnote{In fact, it has been observed that there is differential rotation in the Sun \citep[e.g.,][]{Schou1998ApJ} and some low-mass stars \citep[e.g.,][]{Benomar2018Sci}, but we can make a similar discussion if we regard $\Omega_\mathrm{spin}$ as the averaged angular velocity.}. In equation (\ref{eq:AM_assume2}) both star and planet are assumed to be rotating around the center of the two-body masses with circular orbits \citep[][see also Section \ref{sec:Observation} for our target selection]{Jackson2008ApJ}. It should be noted that $h$ is mostly carried by the AM of the planetary orbit in the case with $M_\mathrm{p}\ll M_\mathrm{s}$.

In addition to the abovementioned simplifications, we  assume that the orbital rotational axis is aligned with the stellar spin axis \citep{Winn2010ApJ,Albrecht2012ApJ, Albrecht2022PASP, Morgan2024AJ} and that the spin of the planet is synchronized to the orbit. The second assumption is justified because the timescale of synchronization \citep[$\sim 1 \, \mathrm{Myr}$ for a close-in Jovian planet;][]{Guillot1996ApJ} is much shorter than the lifetime of the system and because the AM of planetary spin is negligible compared to the total AM.

The time evolution of $J$ and $h$ is described by the following equations \citep[cf.][]{Benbakoura2019AA}: 
\begin{align}
    \frac{\mathrm{d}J}{ \mathrm{d}t} =\Gamma_\mathrm{tide}-\Gamma_\mathrm{wind} \label{eq:star_AM_eq}
\end{align}
and
\begin{align}
    \frac{\mathrm{d}h}{ \mathrm{d}t} = -\Gamma_\mathrm{tide}, \label{eq:planet_AM_eq}
\end{align}
where $\Gamma_\mathrm{wind} (>0)$ and $\Gamma_\mathrm{tide}$ mean the torques through the magnetic braking (``wind torque'', hereafter) and tidal interaction (``tidal torque'', hereafter), respectively. In this formulation, we set the sign of $\Gamma_\mathrm{tide}$ to plus (minus) when the planet orbit the star inside (outside) the co-rotation radius. Figure \ref{fig:Schematic1}, a schematic picture of our system governed by equations (\ref{eq:star_AM_eq}) and (\ref{eq:planet_AM_eq}), shows the behavior of both wind and tidal torques.

Substituting equations (\ref{eq:AM_assume1}) and (\ref{eq:AM_assume2}) into equations (\ref{eq:star_AM_eq}) and (\ref{eq:planet_AM_eq}), we obtain
\begin{align}
    \frac{\mathrm{d} \Omega_\mathrm{spin}}{ \mathrm{d}t}= \frac{\Gamma_\mathrm{tide}-\Gamma_\mathrm{wind}}{I_\mathrm{s}}  \label{eq:star_Omega_eq}
\end{align}
and
\begin{align}
    \frac{\mathrm{d} \Omega_\mathrm{orb}}{ \mathrm{d}t}= \frac{3 \Omega_\mathrm{orb}^{4/3}\Gamma_\mathrm{tide}}{\mu [ G(M_\mathrm{s} + M_\mathrm{p})]^{2/3}}, \label{eq:planet_Omega_eq}
\end{align}
where $\mu = (M_\mathrm{s}M_\mathrm{p})/(M_\mathrm{s}+M_\mathrm{p})$ is the reduced mass. We note that $I_\mathrm{s}$ is assumed to be constant with time, and thus, we neglect the time variation term.

\subsubsection{Formulation of wind torque} \label{subsubsec:wind_torque}

\citet{Skumanich1972ApJ} discovered a simple relation between stellar rotation and age, $t$, via
\begin{align}
    \Omega_\mathrm{spin} \propto t^{1/p}, \label{eq:Skumanich}
\end{align}
where the power-law index takes a standard value of $p \sim 2$. Later observations confirm this trend \citep[][]{Barnes2007ApJ, Angus2015MNRAS}. Without tidal interaction ($\Gamma_\mathrm{tide}=0$), we obtain $\Gamma_\mathrm{wind} \propto \Omega_\mathrm{spin}^{p+1}$
from equations (\ref{eq:star_Omega_eq}) and (\ref{eq:Skumanich}). This power-law relation is supported by one-dimensional magnetohydrodynamical (1D MHD) numerical simulations for Alfv\'{e}n wave-driven magneto-rotating winds conducted by \citet[][]{Shoda2020ApJ}.

However, recent observations suggest complicated behavior that does not obey equations (\ref{eq:wind_torque}): wind torque is considered to be saturated for fast rotators \citep[][]{El-Badry2022MNRAS,Gossage2023ApJ, Belloni2024AA} and weakened for slow rotators \citep[][]{vanSaders2016Nature,Hall2021NatAs}. The detail of these effects is described in Appendix \ref{app:mb_theory}. In this formulation, we include the former effect but ignore the latter because our observational samples are non-slow rotators that are unaffected by this effect (see Section \ref{subsec:validity})

The saturated wind torque in the fast rotator regime is deviated from $\Gamma_\mathrm{wind} \propto \Omega_\mathrm{spin}^{p+1}$ and can be scaled as $\Gamma_\mathrm{wind} \propto \Omega_\mathrm{spin}$ \citep[e.g.,][see also Appendix \ref{app:mb_theory}]{Matt2012ApJ}. Then, we obtain the wind torque formula as 
\begin{align}
    \Gamma_\mathrm{wind} =  
    \begin{cases}
\alpha_\mathrm{mb} \left( \dfrac{\Omega_\mathrm{spin}/2 \pi}{0.05 \, \mathrm{d}^{-1}} \right)^{p+1} \\
\hspace{35pt} \mathrm{for} \ \Omega \leq \Omega_\mathrm{sat} \\[5mm] 
\alpha_\mathrm{mb} \left( \dfrac{\Omega_\mathrm{sat}/2 \pi}{0.05 \, \mathrm{d}^{-1}} \right)^{p}\left( \dfrac{\Omega_\mathrm{spin}/2 \pi}{0.05 \, \mathrm{d}^{-1}} \right) \\
\hspace{35pt} \mathrm{for} \ \Omega > \Omega_\mathrm{sat}
\end{cases}. \label{eq:wind_torque}
\end{align}
$\Omega_\mathrm{sat}$ is the critical angular velocity that divides the saturated and unsaturated regimes. We adopt $\Omega_\mathrm{sat}$ for the Rossby number,  $\mathrm{Ro}=0.1$ (see Appendix  \ref{app:mb_theory}), which corresponds to the threshold for the saturated magnetic activity of low-mass stars \citep[e.g.,][]{Wright2011ApJ, Matt2015ApJ, See2019ApJ}. $\alpha_\mathrm{mb}$ is the reference value in $\Omega_\mathrm{spin}/2 \pi = 0.05 \, \mathrm{d}^{-1}$, when the Skumanich's law is satisfied regardless of stellar mass \footnote{Previous empirical relations adopt the solar spin rate ($\sim 0.035 \, \mathrm{d}^{-1}$) as the reference value \citep[cf.][]{Gossage2023ApJ}. However, recent papers \citep{Metcalfe2022ApJ, Metcalfe2023ApJ, Saunders2024ApJ} suggest that the solar spin rate is so slow that the magnetic braking is weakened (see text and appendix A).} (see Figure \ref{fig:target_WMB}). 
We assume that $\alpha_\mathrm{mb}$, which generally depends on stellar properties, is determined solely by stellar mass, $\alpha_\mathrm{mb}=\alpha_\mathrm{mb}(M_\mathrm{s})$, and is constant over the evolution in our analysis (see the beginning of Section \ref{sec:Method}).

We adopt the wind torque expression of equation (\ref{eq:wind_torque}) and set $p=2$ for simplicity. This simplification enables us to derive $\alpha_\mathrm{mb}$ by comparing the spin-down model with the observed spin rate in the cluster members whose age is estimated (see Appendix \ref{app:alpha_mb}). 

\subsubsection{Formulation of tidal torque} \label{subsubsec:tide_torque}

In a traditional framework \citep[e.g.,][]{Goldreich1966Icar}, the efficiency of tidal interaction is characterized by the modified (dimensionless) quality factor $Q'$, which is inversely proportional to the phase lag between the tidal potential and the tidal bulge divided by the tidal Love number \citep[see][]{Ogilvie2014ARAA}. The tidal interaction is formulated using $Q'$ as \citep[see, e.g.,][]{MurrayDelmott1999,Benbakoura2019AA} 
\begin{align}
  \Gamma_\mathrm{tide} = \mathrm{sign}(\Omega_\mathrm{orb}-\Omega_\mathrm{spin}) \frac{R_\mathrm{s}^5 \Omega_\mathrm{orb}^4 \mu^2}{GM_\mathrm{s}^2} \left( \frac{9}{4Q'} \right). \label{eq:tide_torque}
\end{align}
Here, $R_\mathrm{s}$ is the stellar radius. We note that $ \mathrm{sign}(\Omega_\mathrm{orb}-\Omega_\mathrm{spin})$ controls the behavior of $\Gamma_\mathrm{tide}$ (cf. Figure \ref{fig:Schematic1}), which is discussed in the text after the equations (\ref{eq:star_AM_eq}) and (\ref{eq:planet_AM_eq}).

Determining the specific form of $Q'$ remains a complex problem. Here, we assume that $Q'$ is approximated by the power of the tidal frequency, $\omega_\mathrm{tide} \equiv 2|\Omega_\mathrm{orb}-\Omega_\mathrm{spin}|$, with a floor value:
\begin{align}
    Q' = \max\left[ Q'_\mathrm{0} \left( \frac{\omega_\mathrm{tide} / 2\pi}{2 \, \mathrm{d^{-1}}} \right)^{-q}, Q'_\mathrm{min} \right], \label{eq:Q_model}
\end{align}
where $Q'_0$ means the reference value for $\omega_\mathrm{tide} / 2\pi= 2 \, \mathrm{d^{-1}}$ and $q$ is the power-law index. We consider that $Q'_0$ depends on stellar parameters and then constant in our analysis. $Q'_\mathrm{min} (>0)$ is a minimum value of $Q'$, which corresponds to the maximum efficiency of the tidal interaction.

The power-law relation in equation (\ref{eq:Q_model}) comes from theoretical suggestion by previous papers dealing with the tide in the radiative zone of low-mass stars induced by internal gravity wave
\citep[][]{Zahn1975AA,Goodman1998ApJ, BarkerOgilvie2010MNRAS, Essick2016ApJ, MaFuller2021ApJ}. We note that, while the equilibrium/non-wave-like tide  \citep[][]{Zahn1966AnAp, Zahn1989AA, Remus2012AA, Mathis2016AA} and the dynamical tide in a convective zone induced by inertial wave \citep{Ogilvie2007ApJ,Ogilvie2013MNRAS, Mathis2015AA} are generally suggested as notable mechanisms, \citet{Barker2020MNRAS} claims that they are subdominant for slowly rotating solar-type stars, which are the main target of this paper (see Sections \ref{subsec:Procedure}). However, even when limited to the tide in the radiative zone, the detailed mechanism is not still well understood (Appendix \ref{app:tide_theory}) and there are large uncertainties in $Q'_0$ and $q$. Then, we treat them as the free parameters to constrain $Q'$.

The setting of the lower bound of $Q'_\mathrm{min}$ in equation (\ref{eq:Q_model}) is required to avoid the divergence of  $\Gamma_\mathrm{tide} (\propto 1/Q')$ for $Q'=0$ (equation \ref{eq:tide_torque}) at the co-rotation radius where $\omega_\mathrm{tide}=0$ for a negative value of $q$ \citep[cf.][]{Penev2018AJ}. A non-zero $Q'_\mathrm{min}$ is also reinforced from an observational point of view because there have been already a number of non-synchronized or eccentric close-in Jovian planets identified \citep[e.g.,][]{Matsumura2008ApJ, Matsumura2010ApJ}; these observations contradict a spurious enhancement in $\Gamma_\mathrm{tide}$ at the co-rotation radius because it makes close-in planets synchronized and circularized. We note that the divergence of $Q'$ at $\omega_\mathrm{tide}=0$ does not cause problematic behavior for positive $q$ because it simply gives $\Gamma_\mathrm{tide} (\propto 1/Q')=0$.

The presence of a lower bound of $Q'$ is also required owing to the fact that a variation in the transit timing has not been detected for close-in planets \citep[e.g.,][]{Ivshina2022ApJS, Adams2024arXiv}, except for WASP-12 b \citep{Yee2020ApJL,Turner2021AJ,Wong2022AJ}. We note that the host star WASP-12 has $\sim 1.4 \, M_\odot$ \citep{Collins2017AJ}, which is larger than the low-mass stars we focus on. 

For the WASP-12 -- WASP-12 b system, $Q'$ is estimated to be about $1.5 \times 10^5$ \citep{Wong2022AJ}. For lower-mass stars, there is any confident detection of orbital decay by the transit-time variation, which implies larger $Q'$ \citep[][]{Patra2020AJ, Maciejewski2018AcA,Maciejewski2020AcA,Maciejewski2022AA, Mannaday2022AJ,Rosario2022AA,Harre2023AA}.
Throughout this paper, we set a conservative value of $Q'_\mathrm{min} = 1 \times 10^5$.  We note that results in the following sections are hardly affected by a choice of $Q'_\mathrm{min}$, provided $10^4 \lesssim Q'_\mathrm{min} \lesssim 10^6$ (see Section \ref{subsec:Procedure}). 

\subsection{Trajectories and forbidden region in phase space} \label{subsec:Concept}

\begin{table}
	\centering
	\caption{Model parameters to determine the trajectory in $\Omega_\mathrm{orb}$ -- $\Omega_\mathrm{spin}$ plane from equations (\ref{eq:star_Omega_eq}) (\ref{eq:planet_Omega_eq}), (\ref{eq:wind_torque}), (\ref{eq:tide_torque}), and (\ref{eq:Q_model}).}
 \label{tab:parameter}
	\begin{tabular}{cll} 
        \hline
        \hline
	\multicolumn{3}{l}{Given Parameter} \\
        \hline
        \hline
	Symbol & Description &   Equation \\
        \hline
        $M_\mathrm{s}$ & Stellar mass &   (\ref{eq:planet_Omega_eq})  (\ref{eq:tide_torque}) \\
        $R_\mathrm{s}$ & Stellar radius &   (\ref{eq:tide_torque}) \\
        $I_\mathrm{s}$ & Stellar moment of inertia &  (\ref{eq:star_Omega_eq}) \\
        $M_\mathrm{p}$ & Planetary mass  &  (\ref{eq:planet_Omega_eq})  (\ref{eq:tide_torque}) \\
        $\alpha_\mathrm{mb}$ & Reference value of $\Gamma_\mathrm{wind}$ for 20-day  &   (\ref{eq:wind_torque})  \\
        & spin period ($\Omega_\mathrm{spin}/2\pi = 0.05 \, \mathrm{d^{-1}}$) & \\
        \hline
        \hline
        \multicolumn{3}{l}{Free Parameter} \\
        \hline
        \hline
        Symbol & Description &   Equation \\
        \hline
        $Q_0'$ & Tidal $Q'$ factor for 0.5-day tidal &  (\ref{eq:Q_model}) \\
        & period ($\omega_\mathrm{tide}/2\pi = 2 \, \mathrm{d^{-1}}$) & \\
        $q$ & Power-law index of $Q'$  &   (\ref{eq:Q_model})  \\
        \hline
	\end{tabular}
\end{table}

\begin{table}[t!]
    \centering
    \caption{The tendency of the AM co-evolution. A sign in the cell indicates that of a formula written at the top of the column.
    }
    \label{tab:tendency}
    \setlength{\tabcolsep}{4pt}
    \begin{tabular}{ccclcc} 
        \hline
	Stage & \multicolumn{2}{c}{Input} & & \multicolumn{2}{c}{Output}  \\ 
        \cline{2-3} \cline{5-6} 
        & $\Omega_\mathrm{spin} - \Omega_\mathrm{orb}$ & $\Gamma_\mathrm{tide} - \Gamma_\mathrm{wind}$ & &	$\mathrm{d} {\Omega}_\mathrm{spin} / \mathrm{d}t$ & $\mathrm{d} {\Omega}_\mathrm{orb} / \mathrm{d}t$  \\ 
	\hline
	1 & $+$ & $-$ & & $-$ & $-$ \\
	2 & $-$ & $-$ & & $-$ & $+$\\
  	3 & $-$ & $+$ & & $+$ & $+$\\
	\hline
	\end{tabular} 
\end{table}

\begin{figure}[t!]
\centering
\plotone{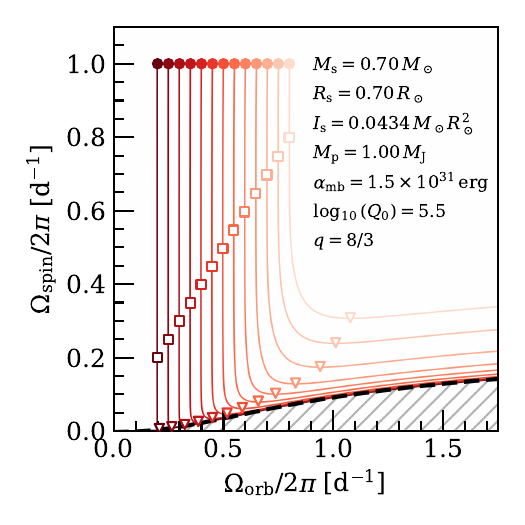}
\caption{Trajectories (colored lines) in phase space of $(\Omega_\mathrm{orb},\Omega_\mathrm{spin})$ derived from equations (\ref{eq:star_Omega_eq}) (\ref{eq:planet_Omega_eq}), (\ref{eq:wind_torque}), (\ref{eq:tide_torque}), and (\ref{eq:Q_model}). We also show the forbidden region (the gray hatched region) and its boundary (the black dashed line), which is obtain by the solution of equation (\ref{eq:trajectory}) that passes the origin. The colors of the lines correspond to the initial values of orbital angular velocity $\Omega_\mathrm{orb,init}$ with the initial condition marked with filled, whereas we set the initial spin rate to a constant value, $\Omega_\mathrm{spin,init}/2 \pi=1 \, \mathrm{d^{-1}}$. The open squares (triangles) represent the transition from stage 1 to stage 2 (from stage 2 to stage 3) in Table \ref{tab:tendency}. We note that other astrophysical settings that are adopted here are written in the upper-right corner (see also texts in Section \ref{subsec:Concept}). It should be mentioned that the boundary is called ``unstable manifold'' in text.
\label{fig:trajectory_sample}}
\end{figure}

\begin{figure}[t!]
\centering
\plotone{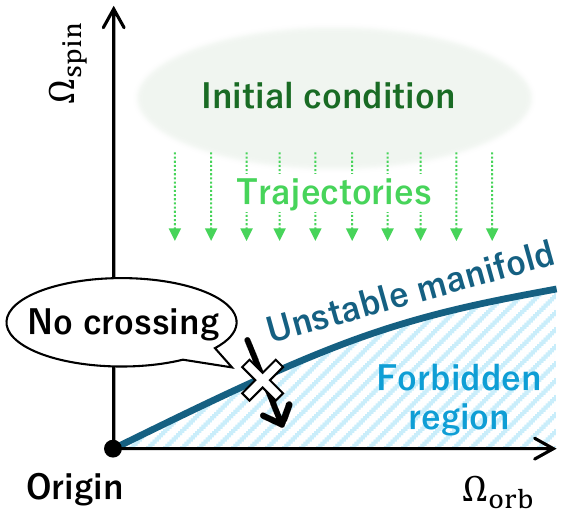}
\caption{Schematic pictures showing the reason for the existence of the forbidden area. The dark green shaded region and light green dotted arrows correspond to the initial condition and trajectories, respectively, which are determined by astrophysical conditions. The dark blue line represents the unstable manifold, which is one of the solutions of equation (\ref{eq:trajectory}) passing through the origin. Because a mathematical argument shows that the unstable manifold can not be intersected (the black arrow with comment), we can show the existence of the forbidden region (the light blue hatched region).  
\label{fig:schematic2}}
\end{figure}

We can track the AM co-evolution of the star-planet system from equations (\ref{eq:star_Omega_eq}) (\ref{eq:planet_Omega_eq}), (\ref{eq:wind_torque}), (\ref{eq:tide_torque}), and (\ref{eq:Q_model}). We describe the characteristic physical properties in the phase space of $(\Omega_\mathrm{orb}, \Omega_\mathrm{spin})$ by using a system composed of a low-mass K-type star with $M_\mathrm{s}=0.70 \, M_\odot$ and $R_\mathrm{s}=0.70 \, R_\odot$ and a Jovian planet with $M_\mathrm{p}=1.0 \, M_\mathrm{J}$ and $R_\mathrm{p}=1.0 \, R_\mathrm{J}$, where $R_\mathrm{p}$ is the planetary radius \footnote{$R_\mathrm{p}$ is used not for tracking the trajectory but for the calculation of the Roche radius. In this paper, we adopt $2.46 \, R_\mathrm{p} (M_\mathrm{s}/M_\mathrm{p})^{1/3}$ as the formulation of Roche radius \citep[cf.][]{CollierCameron2018MNRAS}. }.

We set the initial spin rate of the central star to the range of $\Omega_\mathrm{spin,init}/2\pi \gtrsim  0.1 \, \mathrm{d^{-1}}$, referring to the observed spin rate of stars in ZAMS \citep[ e.g.,][see also Figure \ref{fig:cluster_model}]{Gallet2013AA, Gallet2015AA}. 
We presume that the initial position of the planet is sufficiently outside the Roche radius because otherwise such a system cannot survive for a long time. This condition corresponds to the initial orbital angular velocity $\Omega_\mathrm{orb,init} /2\pi \lesssim 2 \, \mathrm{d^{-1}}$. Except for these two presumptions, there are no restrictions regarding the initial conditions to cover a diversity of planetary systems. 

We calculate the evolution of $(\Omega_\mathrm{orb}, \Omega_\mathrm{spin})$ for standard values of the wind torque parameter, $\alpha_\mathrm{mb}=1.5 \times 10^{31} \, \mathrm{erg}$, reproducing the observed spin-down trend (see Appendix \ref{app:alpha_mb}) and the tidal parameters, $q=8/3$ and $Q'_0= 10^{5.5}$, adopted from the wave-breaking model \citep[][see Appendix \ref{app:tide_theory}]{BarkerOgilvie2010MNRAS,Barker2020MNRAS}, whereas the qualitative behavior described below is preserved regardless of the choice of these parameters. We also assume that the stellar properties, including the moment of inertia, $I_\mathrm{s}=0.0435 \, M_\odot R_\odot^2$, determined from the stellar evolution code \textsc{mesa} \citep[r12778;][see also Appendix \ref{app:MI_MESA}]{Paxton2011APJS, Paxton2013APJS, Paxton2015APJS, Paxton2018APJS, Paxton2019APJS}, are unchanged with time.

Figure \ref{fig:trajectory_sample} shows trajectories in the phase space (the red lines) for the initial stellar spin, $\Omega_\mathrm{spin,init}/2\pi =1.0 \, \mathrm{d^{-1}}$. The trajectories can be separated into three evolutionary stages, listed in Table \ref{tab:tendency}. First, in the stage 1, the stellar spin is decelerated both through the wind torque and the tidal torque because $\Omega_\mathrm{spin} > \Omega_\mathrm{orb}$ (see also Figure \ref{fig:Schematic1}(a)). 
While $\Omega_\mathrm{orb}$ also decreases in the stage 1, it is not explicitly visible in Figure \ref{fig:trajectory_sample} because $\Gamma_\mathrm{wind} \gg \Gamma_\mathrm{tide}$.  Next, in the stage 2, the tidal torque begins to transport AM from the planet to the star as $\Omega_\mathrm{spin} < \Omega_\mathrm{orb}$  (see also Figure \ref{fig:Schematic1}(b)), while a star continues to spin down because $\Gamma_\mathrm{wind} > \Gamma_\mathrm{tide}$. Finally, in the stage 3, the tidal torque overcomes the wind torque, and then both stellar spin and planetary revolution are accelerated. This tendency is summarized in Table 3. 
We note that the observational targets we utilize to constrain the tidal torque are the systems including slowly rotating stars in the stages 2 and 3 listed in Table \ref{tab:tendency}.

A characteristic feature of these trajectories is that they never cross a particular curve (the black dashed line). That is to say, there is the forbidden area (the hatched area) in the phase space of ($\Omega_\mathrm{orb},\Omega_\mathrm{spin}$). The existence and property of the forbidden area are explained by combining astrophysical conditions and mathematical properties of the dynamical system theory. We proceed with the discussion without mathematical rigorous proofs used here, while a brief introduction is presented in Appendix \ref{app:Autonomous} \citep[see, e.g.,][for more detail]{Arnold1992, Hsieh1999}. 

Under the assumption that the stellar parameters are constant, the right-hand side of the governing differential equations (equations \ref{eq:star_Omega_eq} and \ref{eq:planet_Omega_eq})  does not explicitly depend on the independent variable, $t$. Such a system of differential equations is called an 'autonomous system' in the dynamical system theory. When we consider the equations as a two-dimensional autonomous system, the trajectories are obtained as the solution of the tangent of equations (\ref{eq:star_Omega_eq}) and (\ref{eq:planet_Omega_eq}) (see Appendix \ref{app:Autonomous}):
\begin{align}
    \frac{\mathrm{d} \Omega_\mathrm{spin}}{\mathrm{d} \Omega_\mathrm{orb}} = \left( 1-\frac{\Gamma_\mathrm{wind}}{\Gamma_\mathrm{tide}} \right) \frac{\mu [ G(M_\mathrm{s} + M_\mathrm{p})]^{2/3}}{3 \Omega_\mathrm{orb}^{4/3}I_\mathrm{s}}.  \label{eq:trajectory}
\end{align}
Namely, the trajectories on the $\Omega_\mathrm{orb}$ -- $\Omega_\mathrm{spin}$ plane are not explicitly dependent on $t$. Therefore, given the stellar parameters and one passing point, we can track the trajectory through that point. As an important property derived from this, it shows that one trajectory does not intersect with another one (see Appendix \ref{app:Autonomous}).  

Here, we focus on the trajectories passing through the origin, which are unphysical solutions. One trajectory is a trivial solution satisfying $\Omega_\mathrm{orb} = 0$, which is going downward along the $\Omega_\mathrm{spin}$ axis to the origin. The other one, what is noteworthy here, is a non-trivial solution, which appears as a trajectory away from the origin on the phase space (the black dashed line in Figure \ref{eq:trajectory}) \footnote{When a lower bound of $Q'_\mathrm{min}$ is set in equation (\ref{eq:Q_model}), such a solution always exist, even if $q$ is negative.}. In a dynamical system, the latter (former) is called an ``unstable manifold'' (``stable manifold'')  \footnote{The origin corresponds to the saddle equilibrium point in the dynamical system.}. The unstable manifold has two additional characteristic properties besides passing through the origin: 1. the unstable manifold is not intersected by any other solution; 2. both $\Omega_\mathrm{spin}$ and $\Omega_\mathrm{orb}$ are physically required to have non-negative values. According to them, we show that all physically possible trajectories belong only to one of the two subspaces: above or below the unstable manifold. The schematic picture summarizing the discussion is shown in Figure \ref{fig:schematic2}.

The astrophysically reasonable conditions mentioned earlier place the initial position of ($\Omega_\mathrm{orb,init},\Omega_\mathrm{spin,init}$) above the unstable manifold in the phase space, whereas the opposite case will be discussed in Section \ref{subsec:validity}. Thereby, the trajectories for such system can not enter below the unstable manifold, and therefore the forbidden area appears in the phase space of ($\Omega_\mathrm{orb}, \Omega_\mathrm{spin}$).

\subsection{Procedure to constrain tidal quality factor} \label{subsec:Procedure}

\begin{figure}[t!]
\centering
\plotone{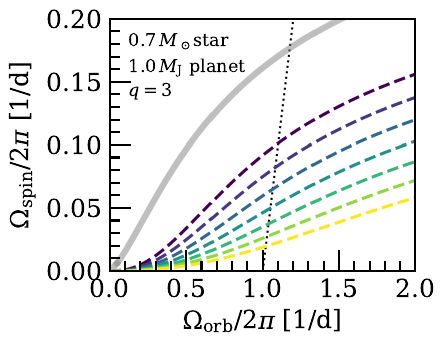}
\plotone{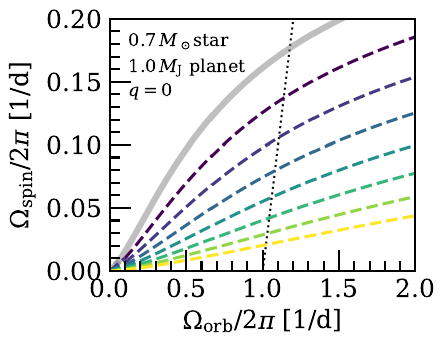}
\plotone{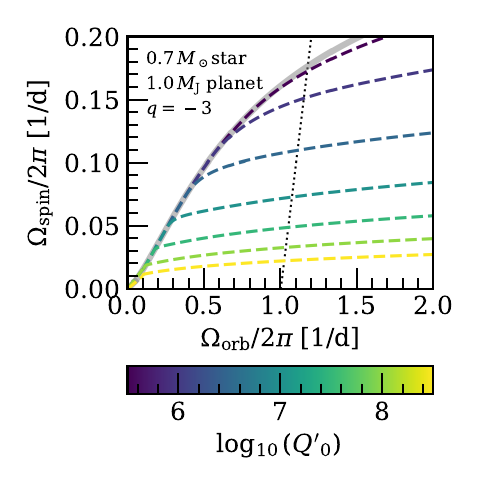}
\caption{The dependence of unstable manifold (color-dotted lines) for a $0.7 \, M_{\odot}$ star $+$ $1 \, M_\mathrm{J}$ planet system on $Q'_0$ when $q=3$ (top panel), $q=0$ (middle panel) and $q=-3$ (bottom panel). The colors of the lines correspond to $Q'_0$ shown in the bottom color-bar. The black dotted lines represent $\omega_\mathrm{tide}/2\pi= 2 \, \mathrm{d^{-1}}$. We also show the unstable manifold for $Q'_0 = Q_\mathrm{min}$ and $q=0$ by the gray solid line. 
\label{fig:unstable_manifold}}
\end{figure}

The trajectories in the $(\Omega_\mathrm{orb},\Omega_\mathrm{spin})$ space depend on the efficiency parameters of the tidal interaction and the magnetic braking in addition to the stellar and planetary parameters (Table \ref{tab:parameter}). Similarly, the solution of the unstable manifold, and accordingly the area occupied by the forbidden region, is dependent on these parameters. Therefore, by comparing observed ($\Omega_\mathrm{orb},\Omega_\mathrm{spin}$) of star-planet systems, we can constrain them; if observed ($\Omega_\mathrm{orb},\Omega_\mathrm{spin}$) was located in the forbidden region, it would require modification of the input value of the parameters. In this paper, we fix the wind torque parameter to the standard value of  $\alpha_\mathrm{mb}=1.5\times 10^{31} \, \mathrm{erg}$ (Section \ref{subsec:Concept} and Appendix \ref{app:alpha_mb}), and give constraints on the tidal quality factor, $Q'$.

Figure \ref{fig:unstable_manifold} shows unstable manifolds for various $Q'_0$ and $q$. They show that smaller $Q'_0$, which causes stronger tidal interaction, results in steeper slopes of the unstable manifold and greater areas of the forbidden region in the phase space. This is because the right-hand side of equation (\ref{eq:trajectory}) ($>0$ for the unstable-manifold solutions) has a negative correlation with $Q'_0$ for any $q$. The dependence on $q$, which looks slightly more complex, can also be understood from the dependence on tidal strength; namely, stronger tidal torque gives a steeper slope of the unstable manifold. In the negative (positive) $q$ cases presented in the bottom (top) panel, the tidal torque is stronger for smaller (larger) $\omega_\mathrm{tide}$ (see equations \ref{eq:tide_torque} and \ref{eq:Q_model}), which corresponds to the upper left (lower right) side of the panel. If one compares the slopes of unstable manifolds for a given $Q'_0$ but with different $q$ values, smaller $q$ cases give steeper (shallower) slopes in the smaller (larger) $\omega_\mathrm{tide}$ side.

In Figure \ref{fig:unstable_manifold} we also plot the unstable manifold for $Q'_0=Q'_\mathrm{min}$ and $q=0$, giving constant $Q'=Q'_\mathrm{min}$  (gray solid line). The forbidden area shaped by this manifold is larger than any other forbidden area formed with $Q'_0>Q'_\mathrm{min}$ for arbitrary $q$.

Let us suppose a situation in which an observed $(\Omega_\mathrm{orb},\Omega_\mathrm{spin})$ in a star-planet system is located in the forbidden region for an arbitrary guessed set of $(q,Q'_0)$. This discrepancy indicates that the initial guess of $Q'_0$ was too small for this $q$ value; we can derive $Q'_\mathrm{0,lim}(>Q'_0)$ that gives the unstable manifold passing exactly through the observed $(\Omega_\mathrm{orb},\Omega_\mathrm{spin})$. The same procedure can be done for different $q$ values so that we can derive the lower bound, $Q'_\mathrm{0,lim}(q)$, as a function of $q$. We note that, as long as the observed point is located inside the maximum forbidden region derived from $ Q'_0 = Q_\mathrm{min}$ and $q=0$ (the gray solid line in Figure \ref{fig:unstable_manifold}), the existence of $Q'_\mathrm{0,lim}(q) \ge Q'_\mathrm{min}$ is guaranteed. 

The novelty of our method is that the unstable manifold does not depend on time and the initial condition of the system. If one can fix the five ``given parameters'' listed in Table \ref{tab:parameter}, a single set of observed $(\Omega_\mathrm{orb},\Omega_\mathrm{spin})$ in a star-planet system at any evolutionary stage can constrain the tidal quality factor in the two-dimensional $(q,Q'_0)$ parameter space. Although this study does not take into account finite eccentricity, $e$, and spin-orbit inclination, $i$, our method can be applied without error as long as $e$ and $i$ are (nearly) zero at the time of observation, even if these took finite values during the evolution before the observation (see the discussion in Section \ref{subsec:non-zero-e-i}).

From an astrophysical point of view, our method gives an upper bound on the tidal torque from a close-in Jovian planet orbiting around a slowly rotating low-mass star. If such s system is observed in the lower right side of the phase space, it will provide a stringent constraint on the tidal torque because sufficiently weak tidal interaction is required to avoid the stellar spin-up by the transfer of AM from the planet. Since this methodology is simple and straightforward, our method is applicable to general tidal torque recipes \citep[e.g.,][]{Barker2020MNRAS} besides the parametrization in equation (\ref{eq:Q_model}).

\section{Target Selection} 
\label{sec:Observation}

\begin{figure}[t!]
\plotone{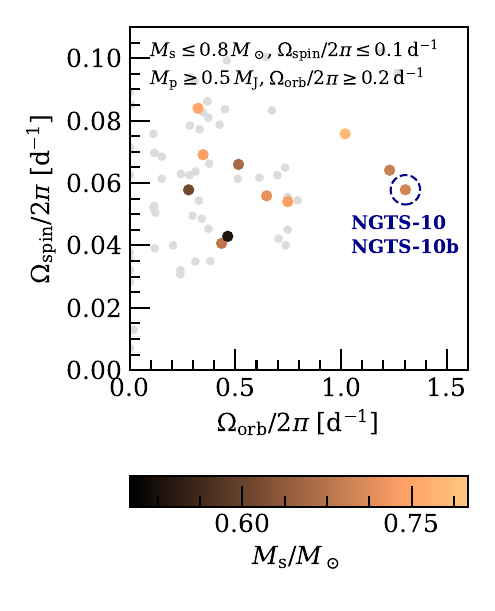}
\caption{Observed values (circle symbols) of the orbital angular velocity of planet $\Omega_\mathrm{orb}$ (abscissa) and the stellar spin rate $\Omega_\mathrm{spin}$ (ordinate), which are obtained from NASA Exoplanet Archive. The large colored symbols show the extracted samples, where the selection criteria for these samples are described in Section \ref{sec:Observation}. The color of the marker refers to the stellar mass $M_\mathrm{s}$ normalized by the Solar mass, which is decoded on the bottom color-bar. The small gray symbols represent the systems with Jovian planets with $M_\mathrm{p} \geq 0.5 \, M_\mathrm{J}$ that do not satisfy the criteria. Our target system, NGTS-10 -- NGTS-10 b, is highlighted.}
\label{fig:observation_NASA}
\end{figure}

\begin{table}[t!]
    \centering
    \caption{Parameters we adopt for the target system, NGTS-10 -- NGTS-10 b. We also show the source of them. }
    \label{tab:parameters_NGTS10}
    \begin{tabular}{lll} 
    \hline
    \hline
    \multicolumn{2}{l}{Object: NGTS-10 -- NGTS-10 b} \\
    \hline
    \hline
    Parameters (star) & Values & Source  \\
    \hline
    $M_\mathrm{s}$ & $0.696 \pm 0.040 \, M_\odot$ & *1\\
    $R_\mathrm{s}$ & $0.697 \pm 0.036 \, R_\odot$ & *1 \\
    $T_\mathrm{eff}$ & $4600 \pm 150 \, \mathrm{K}$ & *1 \\
    $I_\mathrm{s}$ & $0.0435 \, M_\odot R_\odot ^2$ & *2 \\
    $\alpha_\mathrm{mb}$ & $1.5 \times 10^{31} \, \mathrm{erg}$ & *3 \\
    $\Omega_\mathrm{spin}/2\pi$ & $0.0578 \pm 0.0000 \, \mathrm{d^{-1}}$ & *1  \\
    \hline
    Parameters (planet) & Values & Source \\
    \hline
    $M_\mathrm{p}$ & $2.16 \pm 0.10 \, M_\mathrm{J}$ & *1 \\
    $R_\mathrm{p}$ & $1.21 \pm 0.11 \, R_\mathrm{J}$ & *1 \\
    $\Omega_\mathrm{orb}/2\pi$ & $1.3040 \pm 0.000 \, \mathrm{d^{-1}}$ & *1 \\
    \hline
    \end{tabular} 
    \raggedright
    \tablecomments{*1 The literature \citep{McCormac2020MNRAS}, *2 The \textsc{mesa} calculation (see Appendix \ref{app:MI_MESA}), *3 Fitting to the observed spin rates (see Appendix \ref{app:alpha_mb})}
\end{table}

In this section, we briefly introduce the observational data we utilize. Our targets are non-synchronized systems consisting of a slowly rotating low-mass star and a close-in gaseous planet with a circular orbit because their observation can be directly compared with our model (Section \ref{sec:Method}). Then, from the list of observed planetary systems of NASA Exoplanet Archive\footnote{\url{https://exoplanetarchive.ipac.caltech.edu/}} \citep[][]{Akeson2013PASP}, we extract the stars satisfying the following conditions: $M_\mathrm{s} \leq 0.8 \, M_\odot$, $M_\mathrm{p} \geq 0.5 \, M_\mathrm{J}$, $0.02 \leq \Omega_\mathrm{spin}/2\pi \leq 0.1 \, \mathrm{d^{-1}}$, $\Omega_\mathrm{orb}/2\pi \geq 0.2 \, \mathrm{d^{-1}}$, and  $e \leq 0.1$. Figure \ref{fig:observation_NASA} shows the plot of the extracted samples (colored symbols) in the phase space of ($\Omega_\mathrm{orb},\Omega_\mathrm{spin}$) among other samples satisfying $M_\mathrm{p} \geq 0.5 \, M_\mathrm{J}$ (small gray symbols). We can confirm that no data is distributed in the lower-right region of $\Omega_\mathrm{orb}$ -- $\Omega_\mathrm{spin}$ plane; the shape of the empty region looks exactly like a forbidden region in Figures \ref{fig:schematic2} and \ref{fig:unstable_manifold}. 

On the basis of the discussion in Sections \ref{subsec:Concept} and \ref{subsec:Procedure}, the sample with the largest  $\Omega_\mathrm{orb}/\Omega_\mathrm{spin}$ gives the most severe restrictions. Then, we focus on NGTS-10 -- NGTS-10 b \citep{McCormac2020MNRAS}, which was first detected in the Next Generation Transit Survey \citep[NGTS;][]{Wheatley2018MNRAS}, as the suitable target in this paper\footnote{Figure \ref{fig:observation_NASA} shows that WASP-43 system, which is the closest point to the NGTS-10 system in the figure, is also suitable. However, \citet{Davoudi2021AJ} report later that the stellar spin is twice faster than the plotted value \citep[from][]{Espinto2017AA}, which does not satisfy the condition in the statement. Therefore, we eliminate it from the target. }. Table \ref{tab:parameters_NGTS10} shows the pre-determined parameters to constrain the tidal quality factor (see Section \ref{subsec:Procedure}). Most of the stellar and planetary parameters originate from  \citet{McCormac2020MNRAS}, who derived them by multiple techniques including transit photometry, radial velocity measurements, and SED fitting. We ignore observational errors of these quantities. The stellar moment of inertia $I_\mathrm{s}$ is calculated by the gyration radius $\sqrt{I_\mathrm{s}/(M_\mathrm{s}R_\mathrm{s}^2)}$ of the \textsc{mesa} calculation (see Appendix \ref{app:MI_MESA}).  We fix $\alpha_\mathrm{mb}=1.5\times 10^{31} \, \mathrm{erg}$ as already described in Section \ref{subsec:Procedure} (see also Appendix \ref{app:alpha_mb}). 

\section{Constraint of $Q'$ in Target System} \label{sec:Result}

\begin{figure}[t!]
\plotone{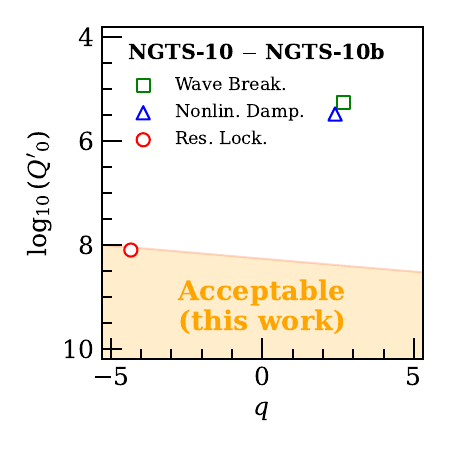}
\plotone{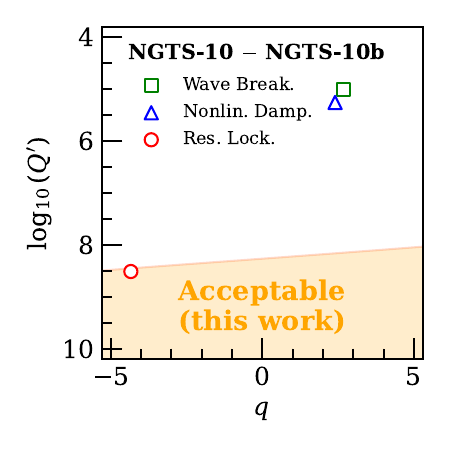}
\caption{Acceptable condition (orange shaded region) in  $(q,Q'_0)$ (upper panel) and $(q,Q')$ (lower panel) for the NGTS-10 system obtained from our method. While abscissae in both panels represent $q$ in equation (\ref{eq:Q_model}), ordinates in the upper and lower panel are $\log_{10}(Q'_0)$ and $\log_{10} (Q')$, respectively. For comparison, we also present theoretically predicted values of $q$ and $Q'_0$ by the wave breaking model \citep[][the green square]{BarkerOgilvie2010MNRAS}, the nonlinear wave damping model \citep[][the blue triangle]{Essick2016ApJ} and the resonance locking model \citep[][the red circle]{MaFuller2021ApJ}.
\label{fig:result_NGTS10}}
\end{figure}

Figure \ref{fig:result_NGTS10} shows the allowed region (orange shade) for the tidal quality factor obtained from the NGTS-10 system. We can derive the boundary, $Q'_\mathrm{0.lim}(q)$, of the allowed region for $Q'_0$ (top panel),  following the procedure described in Section \ref{subsec:Procedure}; we obtain the acceptable range as 
\begin{align}
    \log_{10} \left( Q_{0}' \right) \ge 8.268 + 0.050q. \label{eq:result_Q0}
\end{align}

The constraint on $Q'$ (bottom panel of Figure \ref{fig:result_NGTS10}) is directly obtained from equations (\ref{eq:result_Q0}) by substituting $\omega_\mathrm{tide}/2\pi = 2.48 \, \mathrm{d^{-1}}$ (Table \ref{tab:parameters_NGTS10}) into equation (\ref{eq:Q_model}):
\begin{align}
\label{eq:result_Q} 
\log_{10} \left( Q' \right) &\geq 8.268 + \left[0.050-\log\left(\frac{\omega_\mathrm{tide}/2\pi}{2 \, \mathrm{d^{-1}}} \right) \right] q\\
&= 8.268 + \left[-0.045-\log\left(\frac{\omega_\mathrm{tide}/2\pi}{2.48 \, \mathrm{d^{-1}}} \right) \right] q . 
\notag 
\end{align}
Here, we omit $\max$ in equation (\ref{eq:Q_model}) for simplification (see also Figure \ref{fig:Qmin_dependence}). We note that this is consistent with a statistical result of $\log_\mathrm{10} (Q') \simeq 8.26\pm 0.14$, which was estimated from the relation between the planet/star mass ratio and the orbital separation of more than 200 systems with hot Jupiters, by assuming constant $Q'$ \citep[]{CollierCameron2018MNRAS}. We discuss the $\omega_\mathrm{tide}$ dependency of equation (\ref{eq:result_Q}) in Section \ref{subsec:weak_tide}

Figure \ref{fig:result_NGTS10} also plots previous theoretical results based on the tidal interaction by the wave breaking mechanism \citep[the green square;][]{BarkerOgilvie2010MNRAS}, the non-linear damping mechanism \citep[the blue triangle;][]{Essick2016ApJ}, and the resonance locking \citep[the red circle;][]{MaFuller2021ApJ} (see Appendix \ref{app:tide_theory} for detail). The tidal quality factors derived from the wave braking and non-linear damping mechanisms are more than a hundred times smaller than our lower bound; in other words, these mechanisms predict considerably strong tidal torque, compared with that constrained by the NGTS-10 system. This suggests that the tidal efficiencies of these mechanisms should be weakened in the system with a slow rotator and a close-in gaseous planet. On the other hand, the resonance locking mechanism appears to provide a consistent result with our constraint as this mechanism can quantitatively explain the current position of the NGTS-10 system. However, \citet[][]{MaFuller2021ApJ} show that this mechanism may be unlikely to be effective in the system of a low-mass star and a Jovian planet (see also Appendix \ref{app:tide_theory}). At the moment, we do not know any plausible explanation about the weak tide of this system (see Section \ref{subsec:weak_tide} for detail).

Using the obtained constraint, we can discuss the orbital decay of NGTS-10 b. By substituting equations (\ref{eq:planet_Omega_eq}), (\ref{eq:tide_torque}) and (\ref{eq:result_Q}) and the value of Table \ref{tab:parameters_NGTS10} into the timescale of orbital decay $t_\mathrm{decay} \equiv (\mathrm{d} \ln \Omega_\mathrm{orb}/ \mathrm{d} t)^{-1}$, we derive $t_\mathrm{decay}$ for NGTS-10 b as 
\begin{align}
    t_\mathrm{decay} &= \frac{4Q'}{27} \frac{1}{\Omega_\mathrm{orb}} 
    \left(\frac{M_\mathrm{s}}{M_\mathrm{p}} \right) \left[\frac{G (M_s+M_p)}{ \Omega_\mathrm{orb}^2 R_s^3}  \right]^{5/3} \notag \\
    &\gtrsim 5.6 \times 10^{-0.045q} \, \mathrm{Gyr}. \label{eq:t_decay_NGTS-10}
\end{align}
This large timescale indicates that NGTS-10 b has not significantly migrated inward by the tidal interaction in the past. Also, we can discuss the observability of a variation in the transit timing (see the last two paragraphs of Section \ref{subsubsec:tide_torque}).
The shift of transit arrival time $t_\mathrm{shift}$ during the observational duration $t_\mathrm{dur}$ can be estimated to be $t_\mathrm{shift} = t_\mathrm{dur}^2 / t_\mathrm{decay}$ \citep[][see also \citealt{Birkby2014MNRAS}]{CollierCameron2018MNRAS} \footnote{This expression is obtained from equation (27) in \citet{CollierCameron2018MNRAS}, which is twice as large as that of equation (7) in \citet{Birkby2014MNRAS} .}. Then, from equation (\ref{eq:t_decay_NGTS-10}), we have 
\begin{align}
    t_\mathrm{shift} \lesssim 0.56 \times  10^{0.043q} \times  \left(\frac{t_\mathrm{dur}}{10 \, \mathrm{yr}} \right)^2 \, \mathrm{s}. 
    \label{eq:t_shift_NGTS-10}
\end{align}
We note that this value is about ten times smaller than the previous estimate assuming $Q'=2 \times 10^7$ \citep{McCormac2020MNRAS, Alvarado-Montes2021MNRAS}. Even with the current observational techniques,  $t_\mathrm{shift} \gtrsim 10 \, \mathrm{s}$ is required for detecting the shift \citep[][]{Maciejewski2022AA}. Therefore, equation (\ref{eq:t_shift_NGTS-10}) indicates that the transit timing shift of NGTS-10 b is not detectable within a few decades. That is to say, our method presents a more stringent constraint than what could be obtained from observations of transit timing variations.

\section{Discussion} \label{sec:Discussion}

\subsection{Validity of assumption} \label{subsec:validity}

In our method to constrain the tidal quality factor, we employed several simplified treatments to leverage the mathematical robustness of the unstable manifold (Section \ref{sec:Method}). We here discuss the validity of the simplifications and  associated errors.

\subsubsection{Time-dependent stellar properties}
\label{subsubsec:validity_time_dependency}

The most significant simplification that would affect the results is the assumption that the stellar properties, apart from the rotation rate, remain constant over time. As briefly mentioned in the beginning of Section \ref{sec:Method}, a main sequence star generally expands with evolution as the mean molecular weight in the core increases. We are adopting the observed (current) radius of NGTS-10, which must be larger than the radius averaged over the time from the ZAMS to the current epoch; $R_\mathrm{s}$ in our analysis (Sections \ref{sec:Method}-\ref{sec:Result}) is slightly overestimated.  However, we found that the errors due to this deviation in $R_\mathrm{s}$ is small, because two counteracting effects are canceled each other out. First, the larger $R_\mathrm{s}$ increases the right-hand side of equation (\ref{eq:trajectory}) by strengthening of tidal torque (equation \ref{eq:tide_torque}). This gives a larger forbidden region for fixed $Q'$, and therefore, results in larger $Q'_{0,\mathrm{lim}}$, namely a tighter constraint on tidal torque.  Second, however, larger $I_\mathrm{s}$, as a consequence of the larger $R_\mathrm{s}$ (see Appendix \ref{app:MI_MESA} for detailed calculation), decreases the right-hand side of equation (\ref{eq:trajectory}). This effect, which is obviously opposite to the first one, leads to a looser constraint. Quantitatively, these two effects provide a similar level of errors $\lesssim 0.1$ dex with the opposite signs in equations (\ref{eq:result_Q0}) and (\ref{eq:result_Q}) for stars with $M \leq 0.8 \, M_\odot$, whose radius expands by 5-10 percent. For the target star, NGTS-10, we found that these two effects give comparable errors in absolute magnitude. Hence, the deviation of the constraints in equations (\ref{eq:result_Q0}) and (\ref{eq:result_Q}) is less than 0.1 dex from the constraints given by the treatment using time-dependent $R_\mathrm{s}$ and $I_\mathrm{s}$. In contrast, in more massive stars with $M \geq 0.9 \, M_\odot$, the radius increases more than 20 percent, and hence, the obtained constraints are expected to include an uncertainty of one order of magnitude.

\subsubsection{Wind torque}
\label{subsubsec:validity_wind_torque}

We should also remark on the treatment of the wind torque (Section \ref{subsubsec:wind_torque}); we have assumed the time-independent magnetic braking parameter, $\alpha_\mathrm{mb}$. This treatment is reasonable for the samples in Figure \ref{fig:observation_NASA}, including NGTS-10, because the constant $\alpha_\mathrm{mb}$ well reproduces the trend for stars whose Rossby number is smaller than the solar value (Section \ref{subsubsec:wind_torque}). However, the effect of the weakened magnetic braking \citep{vanSaders2016Nature} should be taken into account for slower rotators (Appendix \ref{app:mb_theory}). As the reduced $\Gamma_\mathrm{wind}$ enlarges the forbidden region (see equation \ref{eq:trajectory}), we should include this effect when we apply our procedure to slow rotators with the Rossby number greater than the solar value (Figure \ref{fig:target_WMB} and Appendix \ref{app:mb_theory}). 

In addition, the bimodality of the spin rate distribution of low mass stars, which is called the ‘intermediate period gap', has been reported as another mysteries in the spin-down evolution \citep{McQuillan2013aMNRAS, McQuillan2014ApJS, Davenport2017ApJ, ReinholdHekker2020AA,Gordon2021ApJ, Lu2022aAJ}. Since this bimodality is particularly evident in K and M-type stars with $T_\mathrm{eff} \lesssim 4500 \, K$, NGTS-10, which has $T_\mathrm{eff} = 4600 \pm 150 \, K$, is marginally subject to this effect. Yet, we consider that this effect is minor because a constant  $\alpha_\mathrm{mb}$ reproduces the spin-rate distribution of cluster members with $T_\mathrm{eff} = 4600 \pm 150 \, K$ reasonably well (Fig.~\ref{fig:cluster_model} in Appendix \ref{app:mb_theory}). In contract, when dealing with the cooler stars where this bimodality is prominent, we should modify the treatment of $\Gamma_\mathrm{wind}$.

\subsubsection{Initial condition}
\label{subsubsec:initial_condition}

Another important assumption in our analysis is that the initial position of ($\Omega_\mathrm{orb,init},\Omega_\mathrm{spin,init}$) needs to be located above the unstable manifold so that it is outside the forbidden region (Figure \ref{fig:schematic2}). A caveat is that the astrophysically reasonable initial condition, $\Omega_\mathrm{spin,init}/2\pi \gtrsim 0.1 \, \mathrm{d}^{-1}$ and $\Omega_\mathrm{orb,init}/2\pi \lesssim 2 \, \mathrm{d}^{-1}$(Section \ref{subsec:Concept}), could break this requirement for strong tide, $Q'_0 \lesssim 10^{6.5}$ (Figure \ref{fig:unstable_manifold}). However, in that case, it is impossible to reach observed  $\Omega_\mathrm{spin}/2\pi \approx 0.05 \, \mathrm{d}^{-1}$, including NGTS-10, (Figure \ref{fig:observation_NASA}) from the reasonable initial value, $\Omega_\mathrm{spin,init}/2\pi \gtrsim 0.1 \, \mathrm{d}^{-1}$ because $\Omega_\mathrm{spin}$ is monotonically increasing along any trajectory below the unstable manifold in the phase space (Figure \ref{fig:unstable_manifold} and  equation \ref{eq:trajectory}) Furthermore, the extremely weak tide expected from the NGTS-10 system (equation \ref{eq:result_Q0}) is safely consistent with the assumption for the initial condition (see also Section \ref{subsec:weak_tide}). We note that the other objects plotted in Figure \ref{fig:observation_NASA} also suggest a weak tide so that the initial condition is well outside the forbidden region; discussion on these objects will be summarized in the forthcoming paper. 

\subsubsection{Beyond Two-body system}

The fundamental presumption of this paper is that we consider the isolated two-body system consisting of a star and a planet (see beginning of Section \ref{sec:Method}).  However, when a third body is present, the evolution of the system will be altered by its gravity; a possible outcome of such three-body interaction is the high-eccentricity migration of a planet in the late evolutionary  stage \citep[cf.][]{Masuda2017AJ}. If the NGTS-10 system had undergone such drastic migration recently, it could be transiently located in the forbidden region of the $\Omega_\mathrm{orb}$ -- $\Omega_\mathrm{spin}$ plane; in this case, we cannot rule out strong tidal torque in the NGTS-10 system.

However, the probability of observing such a system is considered to be low because the timescale of orbital decay, $t_\mathrm{decay}$ (equation \ref{eq:t_decay_NGTS-10}), is generally short in the presence of the strong tidal interaction. If we apply $Q' = 10^6$ as a moderately strong tidal torque to the NGTS-10 system, we obtain $t_\mathrm{decay} \approx 30 \, \mathrm{Myr}$ from equation (\ref{eq:t_decay_NGTS-10}). This timescale is a tiny fraction of the inferred age of NGTS-10 order than $7 \, \mathrm{Gyr}$ \citep{McCormac2020MNRAS}; the probability of a recent three-body interaction in the NGTS-10 system would not be high.

\subsection{Implication from obtained weak tide} \label{subsec:weak_tide}

\begin{figure}[t!]
\plotone{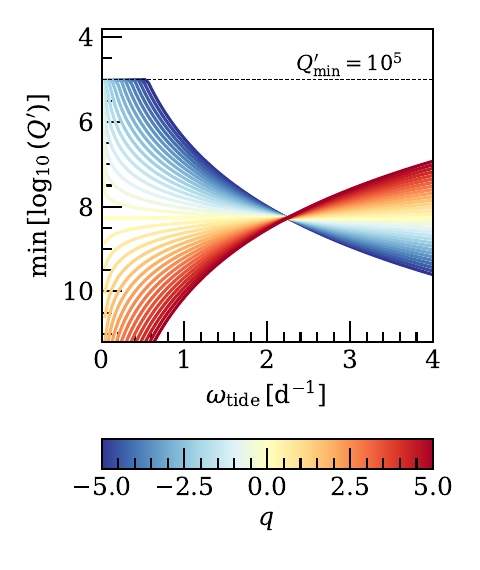}
\caption{Behavior of the lower limit of tidal quality factor obtained from the NGTS-10 system, which is the right-hand side of equation (\ref{eq:result_Q}), as a function of $\omega_\mathrm{tide}$ for various $q$. The colors of lines represent the value of $q$, denoted by the bottom color-bar. We note that $Q'$ is bounded by $Q'\geq Q'_\mathrm{min}=10^5$, which is omitted in equation (\ref{eq:result_Q}). $\omega_\mathrm{tide}=0 \, \mathrm{d^{-1}}$ corresponds to the completely synchronized case. 
\label{fig:Qmin_dependence}}
\end{figure}

From equation (\ref{eq:result_Q}), we obtain the dependency of the lower limit of $Q'$, which corresponds to the right-hand side of equation (\ref{eq:result_Q}), on $\omega_\mathrm{tide}$ and $q$ in Figure \ref{fig:Qmin_dependence}. Supposing that the constraint derived from the NGTS-10 system can be extrapolated to synchronized systems with $\omega_\mathrm{tide}\approx 0$, we verify the tidal properties for different $q$ values in comparison to observed tendencies in synchronized systems on the basis of Figure \ref{fig:Qmin_dependence}.  

Let us begin the discussion with the case of positive $q$, described by redder lines in Figure \ref{fig:Qmin_dependence}. In this case, the tidal torque is suppressed as a system approaches co-rotation ($\omega_\mathrm{tide}=0 \, \mathrm{d^{-1}}$). Then, the tidal interaction in a synchronized system, which satisfies $Q' \gtrsim 10^{8.5}$, is weak enough to be negligible. This tidal properties cannot explain observed synchronized systems with low inclination \citep[e.g.,][see also Section \ref{sec:Introduction}]{Albrecht2012ApJ, Albrecht2022PASP}. In order to solve this inconsistency, we should consider alternative tidal mechanisms, which work for a synchronized system effectively, in addition to the weak tidal interaction with $q>0$ constrained by the NGTS-10 system. For example, the dynamical tide through the inertial wave is considered to be a possible mechanism \citep[e.g.,][]{Mathis2015AA}.

On the other hand, the case of negative $q$, which is described by bluer lines in Figure \ref{fig:Qmin_dependence}, expects a strong tidal interaction in a synchronized system. Therefore, it is not necessary to consider additional effects in this case. Examples of tidal mechanism with negative $q$ include the resonance locking, which satisfies our constraint (Figure \ref{fig:result_NGTS10}). Therefore, it can be a plausible tidal mechanism if further theoretical examination shows that the resonance locking mechanism is also valid for Jovian planets (see also Appendix \ref{app:tide_theory}).

Long-term observations of the shift in transit timing \citep[e.g.,][]{Ivshina2022ApJS, Adams2024arXiv} are a powerful tool to pin down the constraint on the tidal interaction working in synchronized systems. Theoretical investigation with self-consistent numerical calculations \citep[e.g.,][]{Ogilvie2014ARAA} is also essential to elucidate the dominant tidal process operating in various types of two-body systems. 

\subsection{Extension to eccentric and misaligned systems}
\label{subsec:non-zero-e-i}

Our analysis has been restricted on planets in circular orbits on the co-planar plane with the stellar spin. An extension to close-in planets with finite eccentricity, $e$, and spin-orbit inclination angle, $i$, is meaningful because planets with large $e$ or $i$ have been identified with rate of 5-15 percent \citep[][]{Winn2010ApJ, Albrecht2012ApJ, Albrecht2022PASP, Morgan2024AJ}. By employing formulations developed by previous works \citep{Hut1981AA, Eggleton1998ApJ, Ferraz-Mello2008CeMDA, Barker2009MNRAS, Xue2014ApJ}, we can formulate AM co-evolution of eccentric and misaligned planets. We expect to obtain a forbidden region formed by a three dimensional unstable manifold in the four-dimensional phase space of ($\Omega_\mathrm{orb},\Omega_\mathrm{spin},e,i$).
In such a case, we should consider the tidal dissipation not only in the stellar interior but also in the planetary interior, both of which affect the shape of trajectories in the ($\Omega_\mathrm{orb},\Omega_\mathrm{spin},e,i$) space \citep{Jackson2008ApJ,Bonomo2017AA,Efroimsky2022Univ}. Therefore, the shape of the unstable manifolds with respect to the $e$ and $i$ directions is considered to depend on the efficiencies of these two tidal processes. A quantitative investigation of these dependencies is a subject for future work.

It should be noted that the unstable manifolds used in the current paper correspond to the projections of such general cases onto the two-dimensional plane at $e=0$ and $i=0$. Unstable manifolds for finite $e$ and $i$ projected onto that plane are different from the unstable manifold for $e=i=0$. However, from the characteristic properties of the unstable manifold, the system evolves while ensuring that ($\Omega_\mathrm{orb},\Omega_\mathrm{spin}$) remain outside the forbidden region determined by $(e,i)$ at any given time. Therefore, provided that the system has $e=i=0$ at the time of observation, we can safely use the forbidden region in Figure \ref{fig:trajectory_sample} to constrain the tidal torque, regardless of the evolutionary history of $e$ and $i$ (see Section \ref{subsec:Procedure}).

\subsection{Applications to different objects}
\label{subsec:applicability}

Our framework can be directly applied to systems where the spin AM of the companion is negligible (Section \ref{subsec:Framework}). We argued that systems with large $\Omega_\mathrm{orb}/\Omega_\mathrm{spin}$ gives a strong constraint on $Q'$ in Section \ref{subsec:Procedure}. In addition to that, a more massive companion is preferred to provide a tighter constraint because $\mathrm{d} \Omega_\mathrm{spin}/\mathrm{d} \Omega_\mathrm{orb} \propto \mu \approx M_\mathrm{p}$ (equation \ref{eq:trajectory}) so that a larger forbidden region is formed. This is the main reason why we used the massive Jovian planet, NGTS-10 b. 

A promising candidate for such massive companions that readily comes to mind is a brown dwarf. An additional advantage of a brown dwarf around a low-mass star is that its formation is supposed to proceed simultaneously with that of the host star \citep[see, e.g.,][]{Offner2023ASPC}, which enables us to apply the initial conditions for planetary systems (Section \ref{subsec:Concept}) to brown-dwarf systems with a simple extrapolation. We expect the non-synchronized system with a low-mass star and a brown dwarf to provide a tight constraint of $Q'$ even if $\Omega_\mathrm{orb}/\Omega_\mathrm{spin} (>1)$ is relatively small. A caveat is that such systems may be rare firstly because the number of close-in brown dwarfs orbiting K-type stars is depleted compared to brown dwarfs orbiting other types of stars \citep{Grieves2021AA} and secondly because a system with a close-in brown dwarf and an M-type star is easily synchronized owing to their small host-to-companion mass ratio.

We would like to mention the constraint inferred from exoplanetary systems with terrestrial planets because some of them have considerably large $\Omega_\mathrm{orb}/\Omega_\mathrm{spin}$. Unfortunately, however, even the system with the largest $\Omega_\mathrm{orb}/\Omega_\mathrm{spin} \simeq 100$ consisting of an M-type star and a terrestrial planet, KOI-4777 -- KOI-4777 b \citep[][]{Canas2022AJ} \footnote{The mass of KOI-4777 b is not determined \citep{Canas2022AJ}; for this estimate, we assume an Earth mass.}, gives only a weak constraint, $Q'_\mathrm{0} \gtrsim 10^5$, in comparison with the result obtained from the NGTS-10 system (equation \ref{eq:result_Q0}). This is because the advantage of the large $\Omega_\mathrm{orb}/\Omega_\mathrm{spin}$ is overwhelmed by the disadvantage of the small $M_\mathrm{p}$. 

Compact stars, such as white dwarfs and neutron stars, are are also good candidates. Although we have to modify the formulation to consider different formation paths to these objects and relativistic effects particularly for neutron stars, the fundamental concept regarding a forbidden region can be universally applied. However, we should note that it is difficult to observe a non-synchronized system because most of close-in systems frequently observed as cataclysmic stars or X-ray binaries are considered to be synchronized ones \citep[e.g.,][]{Mazeh2008EAS}.

We can extend our framework to the non-synchronized binary consisting of two low-mass stars. For such a system, we should explicitly calculate the evolution of the rotation, $\Omega_\mathrm{spin,2}$, of the secondary star with magnetic braking, in addition to $\Omega_\mathrm{spin,1}$ of the primary star in the governing equations \citep[e.g.,][]{Fleming2019ApJ}. We can follow the same procedure using the two-dimensional unstable manifold in the three-dimensional phase space of ($\Omega_\mathrm{orb},\Omega_\mathrm{spin,1},\Omega_\mathrm{spin,2}$) (cf. Section \ref{subsec:non-zero-e-i}). Indeed, \citet{Lurie2017AJ} reported several such systems in their compilation of eclipsing binaries detected by \textit{Kepler}.

\subsection{Magnetic star-planet interaction} \label{subsec:magnetic_interaction}
It is considered that  magnetic star-planet interaction also plays an important role in the evolution of the system \citep{Cuntz2000ApJ}. If a planetary orbit is located inside the Alfv\'{e}n surface the sub-Alfv\'{e}nic region of the stellar wind, the energy and angular momentum are transferred by the Alfv\'{e}n wings scenario \citep{Cohen2010ApJ,Strugarek2014ApJ}, which is motivated by planet–satellite magnetic interactions in the solar system \citep{GoldreichLyndenbell1969ApJ, Neubauer1980JGR, Neubauer1998JGRN}. The torque produced by the magnetic interaction, $\Gamma_\mathrm{mag}$, is formulated by numerical simulations  \citep{Strugarek2016ApJ,Strugarek2017ApJ}.

By formally replacing $\Gamma_\mathrm{tide}$ with $\Gamma_\mathrm{tide}+\Gamma_\mathrm{mag}$ in equation (\ref{eq:trajectory}), we can extend our formalism. \citet{Ahuir2021AA} conducted comprehensive numerical simulations for the AM co-evolution in the star-planet system, including all $\Gamma_\mathrm{wind}, \Gamma_\mathrm{tide}$ , and $\Gamma_\mathrm{mag}$, and pointed out the forbidden region in the phase space of ($\Omega_\mathrm{orb}, \Omega_\mathrm{spin}$).

\section{Summary and Conclusion} \label{sec:Conclusion}
The aim of this paper is to construct a framework to obtain a robust constraint of the tidal $Q'$ factor that is not subject to the indeterminacy of initial conditions and the uncertainty of stellar ages.  For this purpose, we presented the method to restrict $Q'$ from a single set of observed orbital rotation of a companion object and spin of a host object, ($\Omega_\mathrm{orb}, \Omega_\mathrm{spin}$), by utilizing the argument of dynamical system theory.  

We first introduced the model to calculate the evolution of the AM of stellar spin and the AM of the orbital rotation of a planet under the time-independent stellar and planetary properties. The co-evolution of these AMs is governed by the tidal interaction between the star and the planet and the magnetic braking of the star through stellar winds. By tracking the AM co-evolution from the astrophysically reasonable initial condition, we found that a forbidden region appears in the lower-right part of the $\Omega_\mathrm{orb}$ -- $\Omega_\mathrm{spin}$ plane, regardless of its evolutionary history. On the basis of the dynamical system theory, we also presented that the forbidden region is formed by the unstable manifold that corresponds to the evolutionary track starting from the origin, $(\Omega_\mathrm{orb}, \Omega_\mathrm{spin}) = (0,0)$. As the location and configuration of the unstable manifold are characterized by the strength of the tidal interaction, we proposed that a constraint of $Q'$ is obtained by comparing the observed ($\Omega_\mathrm{orb}, \Omega_\mathrm{spin}$) of a single star-planet system with the forbidden region for given stellar and planetary parameters and magnetic-braking strength. 

Our framework can provide a tight constraint on tidal interaction in a system composed of a slow rotator and a close-in Jovian planet. We demonstrated that NGTS-10 -- NGTS-10 b, the system with the currently known largest $\Omega_\mathrm{orb}/\Omega_\mathrm{spin}$, only allows extremely weak tide , $Q' \gtrsim 10^8$. 
This constraint is in conflict with the $Q'$ values predicted by the wave braking mechanism or the nonlinear-damping mechanism. Among the currently known mechanisms, only the resonance locking is consistent with this constraint, although this mechanism may not be effectively working in Jovian planets \citep{MaFuller2021ApJ}. Additionally, our constraint predicts that the shift of transit timing due to the tidal interaction for this system is too small to detect with the current observational accuracy within a few decades.

Our result obtained from the NGTS-10 system required re-examination of the tidal interaction at least for non-synchronized systems. Also, our proposed framework has the potential to apply to various non-synchronized binary systems including low-mass stars.


\section*{Acknowledgement}

T.T. is supported by IGPEES, WINGS Program in the University of Tokyo and Research Fellowships for Young Scientists in JSPS (24KJ0605). T.K.S. is supported by Grants-in-Aid for Scientific Research from the MEXT/JSPS of Japan, 22H01263.





\appendix

\section{$\Gamma_\mathrm{wind}$ Formalism} \label{app:mb_theory}

\begin{figure}[t!]
\plotone{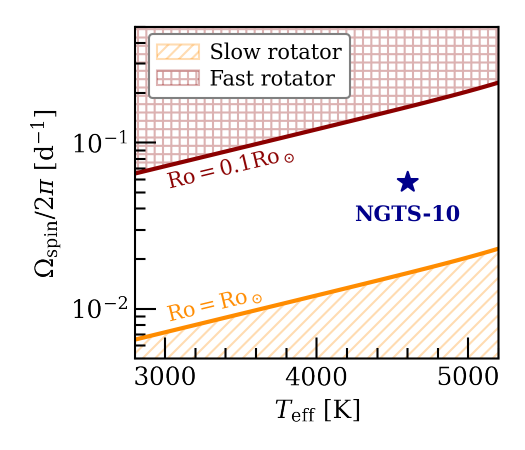}
\caption{Fast-rotator ($\mathrm{Ro}/\mathrm{Ro}_\odot \leq 0.1$; red cross-hatched region) and slow-rotator ($\mathrm{Ro}/\mathrm{Ro}_\odot  \geq 1$; orange diagonal-hatched region) regimes with the target star, NGTS-10 (star symbol), in $T_\mathrm{eff}$ (abscissa) - $\Omega_\mathrm{spin}$ (ordinate) diagram. In order to calculate the Rossby number (equation \ref{eq:Rossbynum}), we adopt the model of \citet{Cranmer2011ApJ} for $\tau_\mathrm{cz} (T_\mathrm{eff})$. 
\label{fig:target_WMB}}
\end{figure}

Theoretical models of magneto-rotator winds \citep[e.g.,][]{WeberDavis1967ApJ, Sakurai1985AA} 
give an approximated analytical expression for the wind torque, 
\begin{align}
  \Gamma_\mathrm{wind} \approx \dot{M} r_\mathrm{A} \Omega_\mathrm{spin}. \label{eq:WD67}
\end{align}
where $\dot{M}$ is the mass-loss rate by the wind, and $r_\mathrm{A}$ expresses the Alfv\'{e}n radius, which is the cylindrical radius where the radial components of the wind and local Alfv\'en velocities are equal. Given that $\dot{M}$ and $r_\mathrm{A}$ depend on stellar parameters and rotational properties \citep[e.g.,][]{MattPudritz2008ApJ, Cranmer2011ApJ, Matt2012ApJ, Suzuki2013PASJ, Shoda2020ApJ}, we can obtain the power-law relation between $\Gamma_\mathrm{wind}$ and $\Omega_\mathrm{spin}$ (equation \ref{eq:wind_torque}), leading to Skumanich's law (equation  \ref{eq:Skumanich}).

Stellar rotational and magnetic activity can be characterized by Rossby number, 
\begin{align}
    \mathrm{Ro} \equiv \frac{2\pi}{\Omega_\mathrm{spin}\tau_\mathrm{cz}}, \label{eq:Rossbynum}
\end{align}
where $\tau_\mathrm{cz}$ is the convective turnover time \citep{Noyes1984ApJ}. 
Fast rotators generally correspond to young magnetically active stars. In this regime, $\Gamma_\mathrm{wind}$ does not rapidly increase with increasing $\Omega_\mathrm{spin}$ but is shifted to the slow increase, $\Gamma_\mathrm{wind}\propto \Omega_\mathrm{spin}$, for $\mathrm{Ro}/\mathrm{Ro}_\odot \lesssim 0.1$ \citep[equation \ref{eq:wind_torque};][]{El-Badry2022MNRAS,Gossage2023ApJ,Belloni2024AA}, in accordance with the saturation of magnetic activity \citep{Wright2011ApJ,Wheatley2018MNRAS,See2019ApJ}. In the slow-rotator regime, $\mathrm{Ro}/\mathrm{Ro}_\odot \gtrsim 1$, $\Gamma_\mathrm{wind}$ is more weakened than expected from the simple power-law dependence \citep{vanSaders2016Nature,Hall2021NatAs,David2022ApJ,Metcalfe2022ApJ,Saunders2024ApJ}, which has also been investigated theoretically \citep{MetcalfevanSaders2017SoPh,Tokuno2023MNRAS}. We do not consider the latter effect because our target star, NGTS-10, is still before entering the slow-rotator regime (Figure \ref{fig:target_WMB}).

\section{Estimation of $\alpha_\mathrm{mb}$} \label{app:alpha_mb}

\begin{table}[t!]
    \centering
    \caption{The list of clusters and their age we use.}
    \label{tab:list_cluster}
    \setlength{\tabcolsep}{4pt}
    \begin{tabular}{lccc} 
		\hline
		Cluster & Age & Ref. (Age) & Ref. ($\Omega_\mathrm{spin}$ and $T_\mathrm{eff}$) \\
       & (Myr) \\
		\hline
            $\alpha$ Persei & 80 & 1 & 1 \\
            Pleiades & 125 & 2 & 2,3 \\
            M50 & 150 & 2 & 2,4 \\
            NGC 2516 & 150 & 2 & 2,5 \\
            Group X & 300 & 6 & 6,7 \\
            NGC 3532 & 300 & 8 & 8 \\
            NGC 2281 & 450 & 9 & 9 \\
            M37  & 500 & 2 & 2,10 \\
            Preasepe & 700 & 2 & 2,11 \\
            NGC 6811 & 950 & 2 & 2,12,13,14,15 \\
            NGC 6819 & 2500 & 16 & 16,17 \\
            Ruprecht 147 & 2700 & 17 & 17,18 \\
            M67 & 4000 & 19 & 19,20 \\
		\hline
    \end{tabular} 
    \raggedright
    \tablerefs{ 1. \citet{Boyle2023AJ}, 2. \citet{Godoy-Rivera2021ApJS}, 3. \citet{Rebull2016AJ}, 4. \citet{Irwin2009MNRAS}, 5. \citet{Irwin2007MNRAS}, 6. \citet{Messina2022AA}, 7. \citet{Newton2022AJ}, 8. \citet[][]{Fritzewski2021AA}, 9. \citet{Fritzewski2023AA}, 10. \citet{Hartman2009ApJ}, 11. \citet[][]{Rebull2017ApJ}, 12. \citet{Meibom2011ApJ}, 13. \citet{Curtis2019ApJ}, 14. \citet{Santos2019ApJS}, 15. \citet{Santos2021ApJS}, 16. \citet[][]{Meibom2015Natur}, 17. \citet{Curtis2020ApJ}, 18. \citet[][]{GrunerBarnes2020AA}, 19. \citet{Gruner2023AA}, 20. \citet[][]{Dungee2022ApJ}. }
\end{table}

\begin{figure}[t!]
\plotone{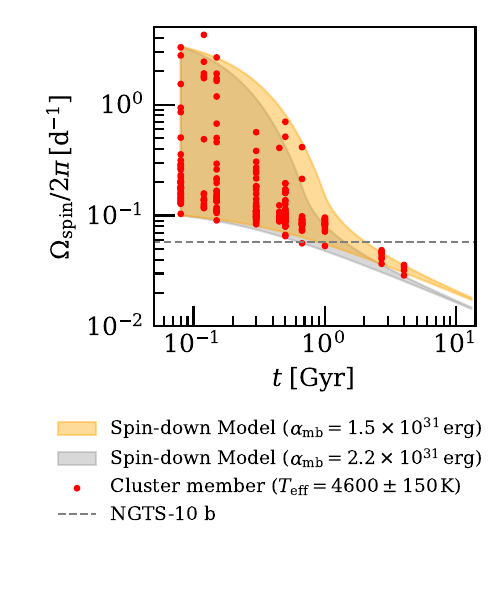}
\caption{Comparison between observed spin rates, $\Omega_\mathrm{spin}/2\pi$, of low-mass stars with $T_\mathrm{eff} = 4600 \pm 150 \, \mathrm{K}$ in various open clusters (red circles) and the spin-down model result for $M_\mathrm{s} = 0.7 \,M_\odot$ and $\alpha_\mathrm{mb} =1.5 \times 10^{31} \, \mathrm{erg}$ (orange shaded region; our estimation) and $2.2 \times 10^{31} \, \mathrm{erg}$ \citep[grey shaded region;][]{Matt2015ApJ} against stellar age, $t$. We set the initial spin rates to cover the distribution of the stars in $\alpha$ Persei at $t=80$ Myr. We also show the spin rate of NGTS-10 as the dashed horizontal line. \label{fig:cluster_model}}
\end{figure}

We are adopting $\alpha_\mathrm{mb}$ $=1.5\times 10^{31} \, \mathrm{erg}$ as the reference value of the wind torque when $\Omega_\mathrm{spin}/2\pi=0.05 \, \mathrm{d^{-1}}$ for the target star, NGTS-10, which has $M_\mathrm{s}=0.70 \, M_{\odot}$. This value is determined to reproduce the observed spin-down trend of low-mass stars with similar mass . For this purpose, we use stars with $T_\mathrm{eff}=4600\pm 150$ K in open clusters (Table \ref{tab:list_cluster}). The \textsc{mesa} calculation shows the duration of the pre-MS before the ZAMS is $\sim 100$ Myr for the star with $M_\mathrm{s} = 0.70 \, M_\odot$. Then, we select the clusters whose age is $\gtrsim 100$ Myr. 

An advantage of using clusters is that the age of a cluster, which is derived from the isochrone fitting in a HR diagram combined with gyrochronology \citep[e.g.,][]{Barnes2003ApJ,Soderblom2010ARAA}, can be used as the reasonable estimate for the stellar age of the cluster members because they are considered to form simultaneously \citep[e.g.,][]{Soderblom2010ARAA} \footnote{Some open clusters show multiple age populations \citep{Cordoni2018ApJ}, which may cause an error in the estimate of the stellar age.}. Stellar spin rates can be measured from the periodic variation of light curves due to star spots \citep[e.g.][]{Garcia2014AA,McQuillan2014ApJS}. 
\textit{Gaia} \citep{Gaia2018AA, Gaia2023AA} provides the color magnitude to derive the effective temperature through the color-$T_\mathrm{eff}$ relation \citep[][]{Curtis2020ApJ} \footnote{We assume the solar metallicity, which is roughly satisfied for our selected samples.}.

We calculate equation (\ref{eq:star_Omega_eq}) with switching off tidal interaction, $\Gamma_\mathrm{tide}=0$, to derive the time evolution of $\Omega_\mathrm{spin}$ from the initial values that encompass the range of the observed stellar spin rates, $0.1 \, \mathrm{d}^{-1} \leq \Omega_\mathrm{spin,init}/2\pi \leq 3.3 \, \mathrm{d}^{-1}$, in $\alpha$ Persei, the youngest cluster with $80$ Myr among our samples. We vary $\alpha_\mathrm{mb}$ but fix the other parameters in Table \ref{tab:parameters_NGTS10} to determine  the best-fit value of $\alpha_\mathrm{mb}=1.5 \times 10^{31} \, \mathrm{erg}$  such that the model calculations with different $\Omega_\mathrm{spin,init}$ cover the lower and upper envelopes of the observed distribution in the $t-\Omega_\mathrm{spin}$ diagram (orange shaded region and red circles in Figure \ref{fig:cluster_model}). This chosen value is moderately smaller than the value, $\alpha_\mathrm{mb}=2.2 \times 10^{31} \, \mathrm{erg}$ (gray shaded region), estimated from the scaling relation calibrated by the solar rotation \citep{Matt2015ApJ}. We consider that the deviation is likely due to newly obtained constraints from the cluster data with ages $\gtrsim$ 2.5 Gyr \citep[][see also Figure \ref{fig:cluster_model}]{Curtis2020ApJ, GrunerBarnes2020AA, Dungee2022ApJ,Gruner2023AA}. 

\section{Theoretical prediction of $Q'$ formulation} \label{app:tide_theory}

In this paper we did not specify the detailed physical mechanism of tidal interaction but used the general expression of the tidal quality factor, equation (\ref{eq:Q_model}). We gave the general constraint, in comparison with the previous results based on the three different mechanisms, in Figure \ref{fig:result_NGTS10}. We briefly introduce these mechanisms with the specific expressions regarding $Q'$ below.

The first process is the wave braking mechanism \citep[][]{BarkerOgilvie2010MNRAS,BarkerOgilvie2011MNRAS}:Large-amplitude internal gravity waves excite nonlinear oscillations, which damp the waves themselves and causes strong dissipation of the wave energy in the radiative core of a low-mass star. \citet[][]{BarkerOgilvie2010MNRAS} obtained
\begin{align}
    Q'_\mathrm{BO10} \simeq 1 \times 10^5 \times  \left( \frac{\mathcal{G}_\mathrm{s}}{ \mathcal{G}_\odot} \right)^{-1} \left( \frac{M_\mathrm{s}}{M_\odot} \right)^{2} \notag \\ 
    \times \left( \frac{R_\mathrm{s}}{R_\odot} \right) \left( \frac{\omega_\mathrm{tide}/2\pi}{2.0 \, \mathrm{d^{-1}}} \right)^{-8/3}. \label{eq:Q_BO10}
\end{align}
from their numerical simulations, where $\mathcal{G}_\mathrm{s}$ is a parameter that depends on the stellar structure. Equation (\ref{eq:Q_BO10}) gives $Q'_0=1.8 \times 10^5$ and $q=8/3$ for NGTS-10 (green square in Figure \ref{fig:result_NGTS10}) , which yield $Q'=1.0 \times 10^5$ listed in \citet[][]{Barker2020MNRAS}.

The second process is the nonlinear wave damping mechanisms: the numerical simulation conducted by \citet{Essick2016ApJ} shows the nonlinear oscillation itself leads to energy dissipation through cascades, even in the absence of the wave breaking, in the system composed of a low-mass star and a hot Jupiter. They showed the tidal quality factor that is available for Jovian planets with $M_\mathrm{p} \gtrsim 0.3 \, M_\mathrm{J}$ as 
\begin{align}
    Q'_\mathrm{EW16} \simeq 2 \times 10^5 \times \left( \frac{M_\mathrm{p}}{M_\mathrm{J}} \right) ^{1/2} 
    \left( \frac{\omega_\mathrm{tide}/2\pi}{2.0 \, \mathrm{d^{-1}}} \right)^{-2.4},
    \label{eq:Q_EW16}
\end{align}
providing $Q'_0=3.0 \times 10^5$ and $q=2.4$ for NGTS-10 (blue circle in Figure \ref{fig:result_NGTS10}).

The third process is the resonance locking mechanism,  suggested by \citet{Fuller2017MNRAS}. The energy dissipation by the tidal force is enhanced by the resonance between the tidal frequency and the stellar g-mode frequency. \citet{MaFuller2021ApJ} investigated the efficiency of this effect for a low-mass star and derived the tidal quality factor as 
\begin{align}
    Q'_\mathrm{MF21} \simeq 2 \times 10^6 \times \left( \frac{M_\mathrm{p}}{M_\mathrm{J}} \right) \left( \frac{M_\mathrm{s}}{M_\odot} \right)^{-8/3} \left( \frac{R_\mathrm{s}}{R_\odot} \right)^5 \notag \\ 
    \times \left( \frac{t_\alpha}{5 \, \mathrm{Gyr}} \right) \left( \frac{\Omega_\mathrm{orb}/2\pi}{0.5 \, \mathrm{d^{-1}}} \right)^{13/3},
    \label{eq:Q_MF21}
\end{align}
where $t_\alpha$ is the mode evolution timescale, which is determined by the stellar structure and evolution. We use $\Omega_\mathrm{orb} \gg \Omega_\mathrm{spin}$, namely, $\Omega_\mathrm{orb} \approx \omega_\mathrm{tide}/2$, when comparing with equation (\ref{eq:Q_model}).
It should be mentioned that the resonance locking may operate only for low-mass planets with $M_\mathrm{p} \lesssim 0.1 \, M_\mathrm{J}$ because the non-linear effects prevent the resonance \citep[see][]{MaFuller2021ApJ}. In this paper, however, we assume this mechanism occurs on NGTS-10 b to give $Q'_0=1.3 \times 10^8$ and $q=-13/3$ (red circle in Figure \ref{fig:result_NGTS10})\footnote{In \citet{MaFuller2021ApJ}, $Q'_0$ for NGTS-10 was not explicitly presented. We used the stellar parameters for WASP-43, a star similar to NGTS-10, presented in their paper, to derive these values.}.  

\section{Calculation of $I_\mathrm{s}$} \label{app:MI_MESA}

\begin{table}[t!]
    \centering
    \caption{The setting of the \textsc{mesa} stellar evolution code. }
    \label{tab:mesa_setting}
    \begin{tabular}{llc} 
		\hline
		Control name & Our setting & Ref. \\ 
		\hline
		Initial abundance & The solar metallicity & 1 \\
		Reaction net work & ``pp\_and\_cno\_extras.net'' & \\
		Reaction rates & JINA REACLIB & 2 \\
		Atmosphere & ``Eddington'' & \\
		Mixing length & $\alpha_\mathrm{MLT}=2.0$ &- \\
		Diffusion & No & \\
		Overshooting & No & \\
  Mass Loss & No & \\
		\hline
    \end{tabular} 
    \raggedright
    \tablerefs{1. \citet{GrevesseSauval1998SSRv}, 2. \citet{Cyburt2010ApJS}}
\end{table}

\begin{figure}[t!]
\plotone{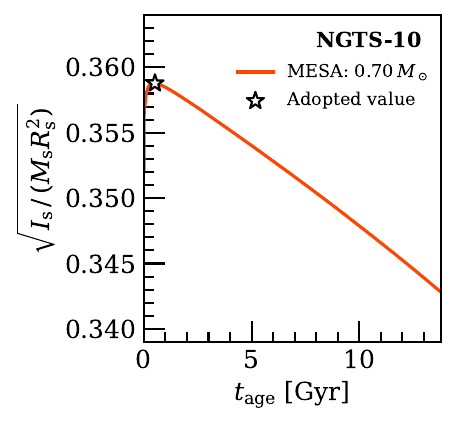}
\caption{Time evolution of $\hat{r}_\mathrm{g}$ for $M_\mathrm{s}=0.7M_\odot$ computed by the \textsc{mesa} (the orange line). The star-shaped symbol represents the adopted value for the time-independent $I_\mathrm{s}$ in our calculations (see text for detail).
\label{fig:MESA_NGTS-10}}
\end{figure}

We determine $I_\mathrm{s}$, which is assumed to be time-independent (Section \ref{sec:Method}), by using \textsc{mesa} stellar evolution code (r12278) with the input parameters summarized in Table \ref{tab:mesa_setting}\footnote{This setting reproduces the physical properties of the present Sun well. We upload the inlists in Zenodo: \dataset[10.5281/zenodo.8385153]{https://doi.org/10.5281/zenodo.11441698} }. To follow the evolution of NGTS-10, we calculate the time evolution of a $M_\mathrm{s}=0.7M_{\odot}$ star from the ZAMS to 14 Gyr ($\approx$ the age of the universe), which is before the end of the MS phase. Figure \ref{fig:MESA_NGTS-10} presents the time variation of a dimensionless quantity, gyration radius, $\hat{r}_\mathrm{g} \equiv\sqrt{I_\mathrm{s}/(M_\mathrm{s}R_\mathrm{s}^2)}$ \citep[cf.][]{Damiani2015AA,Damiani2016AA}. $\hat{R}_\mathrm{g}$ slowly decreases with time as the star evolves to core-halo-like structure. We are employing 
\begin{align}
I_\mathrm{s}\equiv M_\mathrm{s}R_\mathrm{s}^2 \hat{r}_\mathrm{g,max}^2
\label{eq:Is}
\end{align}
for our calculations, where $\hat{r}_\mathrm{g,max}^2$ is the maximum gyration radius shown by the star symbol in Figure \ref{fig:MESA_NGTS-10}. 

Equation (\ref{eq:Is}) obviously overestimates $I_\mathrm{s}$ firstly because $\hat{r}_\mathrm{g,max}$ is larger than $\hat{r}_\mathrm{g}$ averaged over the evolution until the current time and secondly because $R_\mathrm{s}$ adopted from the observed value (Table \ref{tab:parameters_NGTS10}) is also larger than the time averaged radius (Section \ref{subsec:validity}). By this  overestimate of $I_\mathrm{s}$ we can avoid overconstraining the tidal quality factor, as discussed in Section \ref{subsec:validity}. The deviation of the constraint on $Q'$ due to the overestimate is within $\lesssim 0.1$ dex with respect to the NGTS-10 system.

\section{Autonomous system} \label{app:Autonomous}
The theoretical background of this paper is based on the fact that the systems consisting of equations (\ref{eq:star_AM_eq}) and (\ref{eq:planet_AM_eq}) can be treated as two-dimensional autonomous systems under some presumptions. In this section, we briefly introduce the proof of the property of the autonomous system used in Section \ref{subsec:Concept} \citep[e.g.,][for detail]{Arnold1992, Hsieh1999}. 

We consider an $n$-dimensional autonomous system
\begin{align}
    \frac {\mathrm{d} \bm{x}(t)}{ \mathrm{d} t }= \bm{f}\left( \bm{x} \right) \Leftrightarrow \begin{cases}
        \dfrac{\mathrm{d} x_1(t)}{ \mathrm{d} t } &= f_1 \left( x_1, \cdots ,x_n \right) \\[2mm]
        & \vdots \\
        \dfrac{\mathrm{d} x_n(t)}{ \mathrm{d} t } &= f_n \left( x_1, \cdots ,x_n\right) \\[2mm]
    \end{cases}, \label{eq:autonomous}
\end{align}
where $t$, $x_i(t)$, and $f_i$ (for $i = 1, \cdots ,N$) represent the time as the independent variable, the element of the Euclid space as a dependent variable, and the Lipschitz-continous function whose arguments are $x_i$, respectively. For abbreviation, we use the vector notation $\bm{x} = (x_1,\cdots,x_n)^\mathrm{T}$ and $\bm{f} (\bm x) = (f_1 (\bm{x}),\cdots,f_n (\bm{x}))^\mathrm{T}$, where ``T'' means transposition of a vector. At this time, the solution of equation (\ref{eq:autonomous}) is shown as the trajectory in the $n$-dimensional phase space of $(x_1, \cdots ,x_n)$. Then, we can deduce that the tangent vector of the trajectory is equal to $\bm{f} (\bm{x})$.

Here, let $\bm{x_*}(t)$ be the unique solution of the initial-value problem for equation (\ref{eq:autonomous}) when $\bm{x}(0)=\bm x_\mathrm{init}$, where $\bm{x}_\mathrm{init}$ is a particular vector. In this case, $\bm{x_*}(t-t_0)$ for an arbitrarily selected time $t_0$ solves the initial-value problem for equation (\ref{eq:autonomous}) when $\bm{x}(t_0)=\bm x_\mathrm{init}$. Therefore, we find that the trajectory in the phase space is determined independently of the value of $t$. Moreover, it indicates that two trajectories that pass through the same point are identical, which is equivalent to that two different trajectories have no intersection.






\bibliography{sample631}{}

\begin{thebibliography}{}
\expandafter\ifx\csname natexlab\endcsname\relax\def\natexlab#1{#1}\fi
\providecommand{\url}[1]{\href{#1}{#1}}
\providecommand{\dodoi}[1]{doi:~\href{http://doi.org/#1}{\nolinkurl{#1}}}
\providecommand{\doeprint}[1]{\href{http://ascl.net/#1}{\nolinkurl{http://ascl.net/#1}}}
\providecommand{\doarXiv}[1]{\href{https://arxiv.org/abs/#1}{\nolinkurl{https://arxiv.org/abs/#1}}}

\bibitem[{{Adams} {et~al.}(2024){Adams}, {Jackson}, {Sickafoose}, {Morgenthaler}, {Worters}, {Stubbers}, {Carson}, {Bhure}, {Dekeyser}, {Huang}, \& {Weinberg}}]{Adams2024arXiv}
{Adams}, E.~R., {Jackson}, B., {Sickafoose}, A.~A., {et~al.} 2024, arXiv e-prints, arXiv:2404.07339, \dodoi{10.48550/arXiv.2404.07339}

\bibitem[{{Ahuir} {et~al.}(2021){Ahuir}, {Strugarek}, {Brun}, \& {Mathis}}]{Ahuir2021AA}
{Ahuir}, J., {Strugarek}, A., {Brun}, A.~S., \& {Mathis}, S. 2021, \aap, 650, A126, \dodoi{10.1051/0004-6361/202040173}

\bibitem[{{Akeson} {et~al.}(2013){Akeson}, {Chen}, {Ciardi}, {Crane}, {Good}, {Harbut}, {Jackson}, {Kane}, {Laity}, {Leifer}, {Lynn}, {McElroy}, {Papin}, {Plavchan}, {Ram{\'\i}rez}, {Rey}, {von Braun}, {Wittman}, {Abajian}, {Ali}, {Beichman}, {Beekley}, {Berriman}, {Berukoff}, {Bryden}, {Chan}, {Groom}, {Lau}, {Payne}, {Regelson}, {Saucedo}, {Schmitz}, {Stauffer}, {Wyatt}, \& {Zhang}}]{Akeson2013PASP}
{Akeson}, R.~L., {Chen}, X., {Ciardi}, D., {et~al.} 2013, \pasp, 125, 989, \dodoi{10.1086/672273}

\bibitem[{{Albrecht} {et~al.}(2012){Albrecht}, {Winn}, {Johnson}, {Howard}, {Marcy}, {Butler}, {Arriagada}, {Crane}, {Shectman}, {Thompson}, {Hirano}, {Bakos}, \& {Hartman}}]{Albrecht2012ApJ}
{Albrecht}, S., {Winn}, J.~N., {Johnson}, J.~A., {et~al.} 2012, \apj, 757, 18, \dodoi{10.1088/0004-637X/757/1/18}

\bibitem[{{Albrecht} {et~al.}(2022){Albrecht}, {Dawson}, \& {Winn}}]{Albrecht2022PASP}
{Albrecht}, S.~H., {Dawson}, R.~I., \& {Winn}, J.~N. 2022, \pasp, 134, 082001, \dodoi{10.1088/1538-3873/ac6c09}

\bibitem[{{Alvarado-Montes} {et~al.}(2021){Alvarado-Montes}, {Sucerquia}, {Garc{\'\i}a-Carmona}, {Zuluaga}, {Spitler}, \& {Schwab}}]{Alvarado-Montes2021MNRAS}
{Alvarado-Montes}, J.~A., {Sucerquia}, M., {Garc{\'\i}a-Carmona}, C., {et~al.} 2021, \mnras, 506, 2247, \dodoi{10.1093/mnras/stab1081}

\bibitem[{{Angus} {et~al.}(2015){Angus}, {Aigrain}, {Foreman-Mackey}, \& {McQuillan}}]{Angus2015MNRAS}
{Angus}, R., {Aigrain}, S., {Foreman-Mackey}, D., \& {McQuillan}, A. 2015, \mnras, 450, 1787, \dodoi{10.1093/mnras/stv423}

\bibitem[{{Arnold}(1992)}]{Arnold1992}
{Arnold}, V.~I. 1992, {Ordinary Differential Equations} (Springer Science \& Business Media)

\bibitem[{{Baraffe} {et~al.}(2015){Baraffe}, {Homeier}, {Allard}, \& {Chabrier}}]{Baraffe2015AA}
{Baraffe}, I., {Homeier}, D., {Allard}, F., \& {Chabrier}, G. 2015, \aap, 577, A42, \dodoi{10.1051/0004-6361/201425481}

\bibitem[{{Barker}(2020)}]{Barker2020MNRAS}
{Barker}, A.~J. 2020, \mnras, 498, 2270, \dodoi{10.1093/mnras/staa2405}

\bibitem[{{Barker} \& {Ogilvie}(2009)}]{Barker2009MNRAS}
{Barker}, A.~J., \& {Ogilvie}, G.~I. 2009, \mnras, 395, 2268, \dodoi{10.1111/j.1365-2966.2009.14694.x}

\bibitem[{{Barker} \& {Ogilvie}(2010)}]{BarkerOgilvie2010MNRAS}
---. 2010, \mnras, 404, 1849, \dodoi{10.1111/j.1365-2966.2010.16400.x}

\bibitem[{{Barker} \& {Ogilvie}(2011)}]{BarkerOgilvie2011MNRAS}
---. 2011, \mnras, 417, 745, \dodoi{10.1111/j.1365-2966.2011.19322.x}

\bibitem[{{Barnes}(2003)}]{Barnes2003ApJ}
{Barnes}, S.~A. 2003, \apj, 586, 464, \dodoi{10.1086/367639}

\bibitem[{{Barnes}(2007)}]{Barnes2007ApJ}
---. 2007, \apj, 669, 1167, \dodoi{10.1086/519295}

\bibitem[{{Belloni} {et~al.}(2024){Belloni}, {Schreiber}, {Moe}, {El-Badry}, \& {Shen}}]{Belloni2024AA}
{Belloni}, D., {Schreiber}, M.~R., {Moe}, M., {El-Badry}, K., \& {Shen}, K.~J. 2024, \aap, 682, A33, \dodoi{10.1051/0004-6361/202347931}

\bibitem[{{Benbakoura} {et~al.}(2019){Benbakoura}, {R{\'e}ville}, {Brun}, {Le Poncin-Lafitte}, \& {Mathis}}]{Benbakoura2019AA}
{Benbakoura}, M., {R{\'e}ville}, V., {Brun}, A.~S., {Le Poncin-Lafitte}, C., \& {Mathis}, S. 2019, \aap, 621, A124, \dodoi{10.1051/0004-6361/201833314}

\bibitem[{{Benomar} {et~al.}(2018){Benomar}, {Bazot}, {Nielsen}, {Gizon}, {Sekii}, {Takata}, {Hotta}, {Hanasoge}, {Sreenivasan}, \& {Christensen-Dalsgaard}}]{Benomar2018Sci}
{Benomar}, O., {Bazot}, M., {Nielsen}, M.~B., {et~al.} 2018, Science, 361, 1231, \dodoi{10.1126/science.aao6571}

\bibitem[{{Birkby} {et~al.}(2014){Birkby}, {Cappetta}, {Cruz}, {Koppenhoefer}, {Ivanyuk}, {Mustill}, {Hodgkin}, {Pinfield}, {Sip{\H{o}}cz}, {Kov{\'a}cs}, {Saglia}, {Pavlenko}, {Barrado}, {Bayo}, {Campbell}, {Catalan}, {Fossati}, {G{\'a}lvez-Ortiz}, {Kenworthy}, {Lillo-Box}, {Mart{\'\i}n}, {Mislis}, {de Mooij}, {Nefs}, {Snellen}, {Stoev}, {Zendejas}, {del Burgo}, {Barnes}, {Goulding}, {Haswell}, {Kuznetsov}, {Lodieu}, {Murgas}, {Palle}, {Solano}, {Steele}, \& {Tata}}]{Birkby2014MNRAS}
{Birkby}, J.~L., {Cappetta}, M., {Cruz}, P., {et~al.} 2014, \mnras, 440, 1470, \dodoi{10.1093/mnras/stu343}

\bibitem[{{Bonomo} {et~al.}(2017){Bonomo}, {Desidera}, {Benatti}, {Borsa}, {Crespi}, {Damasso}, {Lanza}, {Sozzetti}, {Lodato}, {Marzari}, {Boccato}, {Claudi}, {Cosentino}, {Covino}, {Gratton}, {Maggio}, {Micela}, {Molinari}, {Pagano}, {Piotto}, {Poretti}, {Smareglia}, {Affer}, {Biazzo}, {Bignamini}, {Esposito}, {Giacobbe}, {H{\'e}brard}, {Malavolta}, {Maldonado}, {Mancini}, {Martinez Fiorenzano}, {Masiero}, {Nascimbeni}, {Pedani}, {Rainer}, \& {Scandariato}}]{Bonomo2017AA}
{Bonomo}, A.~S., {Desidera}, S., {Benatti}, S., {et~al.} 2017, \aap, 602, A107, \dodoi{10.1051/0004-6361/201629882}

\bibitem[{{Borucki} {et~al.}(2010){Borucki}, {Koch}, {Basri}, {Batalha}, {Brown}, {Caldwell}, {Caldwell}, {Christensen-Dalsgaard}, {Cochran}, {DeVore}, {Dunham}, {Dupree}, {Gautier}, {Geary}, {Gilliland}, {Gould}, {Howell}, {Jenkins}, {Kondo}, {Latham}, {Marcy}, {Meibom}, {Kjeldsen}, {Lissauer}, {Monet}, {Morrison}, {Sasselov}, {Tarter}, {Boss}, {Brownlee}, {Owen}, {Buzasi}, {Charbonneau}, {Doyle}, {Fortney}, {Ford}, {Holman}, {Seager}, {Steffen}, {Welsh}, {Rowe}, {Anderson}, {Buchhave}, {Ciardi}, {Walkowicz}, {Sherry}, {Horch}, {Isaacson}, {Everett}, {Fischer}, {Torres}, {Johnson}, {Endl}, {MacQueen}, {Bryson}, {Dotson}, {Haas}, {Kolodziejczak}, {Van Cleve}, {Chandrasekaran}, {Twicken}, {Quintana}, {Clarke}, {Allen}, {Li}, {Wu}, {Tenenbaum}, {Verner}, {Bruhweiler}, {Barnes}, \& {Prsa}}]{Borucki2010Sci}
{Borucki}, W.~J., {Koch}, D., {Basri}, G., {et~al.} 2010, Science, 327, 977, \dodoi{10.1126/science.1185402}

\bibitem[{{Bouvier} {et~al.}(1997){Bouvier}, {Forestini}, \& {Allain}}]{Bouvier1997AA}
{Bouvier}, J., {Forestini}, M., \& {Allain}, S. 1997, \aap, 326, 1023

\bibitem[{{Boyle} \& {Bouma}(2023)}]{Boyle2023AJ}
{Boyle}, A.~W., \& {Bouma}, L.~G. 2023, \aj, 166, 14, \dodoi{10.3847/1538-3881/acd3e8}

\bibitem[{{Ca{\~n}as} {et~al.}(2022){Ca{\~n}as}, {Mahadevan}, {Cochran}, {Bender}, {Feigelson}, {Harman}, {Kopparapu}, {Caceres}, {Diddams}, {Endl}, {Ford}, {Halverson}, {Hearty}, {Jones}, {Kanodia}, {Lin}, {Metcalf}, {Monson}, {Ninan}, {Ramsey}, {Robertson}, {Roy}, {Schwab}, \& {Stef{\'a}nsson}}]{Canas2022AJ}
{Ca{\~n}as}, C.~I., {Mahadevan}, S., {Cochran}, W.~D., {et~al.} 2022, \aj, 163, 3, \dodoi{10.3847/1538-3881/ac3088}

\bibitem[{{Choi} {et~al.}(2016){Choi}, {Dotter}, {Conroy}, {Cantiello}, {Paxton}, \& {Johnson}}]{Choi2016ApJ}
{Choi}, J., {Dotter}, A., {Conroy}, C., {et~al.} 2016, \apj, 823, 102, \dodoi{10.3847/0004-637X/823/2/102}

\bibitem[{{Cohen} {et~al.}(2010){Cohen}, {Drake}, {Kashyap}, {Sokolov}, \& {Gombosi}}]{Cohen2010ApJ}
{Cohen}, O., {Drake}, J.~J., {Kashyap}, V.~L., {Sokolov}, I.~V., \& {Gombosi}, T.~I. 2010, \apjl, 723, L64, \dodoi{10.1088/2041-8205/723/1/L64}

\bibitem[{{Collier Cameron} \& {Jardine}(2018)}]{CollierCameron2018MNRAS}
{Collier Cameron}, A., \& {Jardine}, M. 2018, \mnras, 476, 2542, \dodoi{10.1093/mnras/sty292}

\bibitem[{{Collins} {et~al.}(2017){Collins}, {Kielkopf}, \& {Stassun}}]{Collins2017AJ}
{Collins}, K.~A., {Kielkopf}, J.~F., \& {Stassun}, K.~G. 2017, \aj, 153, 78, \dodoi{10.3847/1538-3881/153/2/78}

\bibitem[{{Cordoni} {et~al.}(2018){Cordoni}, {Milone}, {Marino}, {Di Criscienzo}, {D'Antona}, {Dotter}, {Lagioia}, \& {Tailo}}]{Cordoni2018ApJ}
{Cordoni}, G., {Milone}, A.~P., {Marino}, A.~F., {et~al.} 2018, \apj, 869, 139, \dodoi{10.3847/1538-4357/aaedc1}

\bibitem[{{Cranmer} \& {Saar}(2011)}]{Cranmer2011ApJ}
{Cranmer}, S.~R., \& {Saar}, S.~H. 2011, \apj, 741, 54, \dodoi{10.1088/0004-637X/741/1/54}

\bibitem[{{Cuntz} {et~al.}(2000){Cuntz}, {Saar}, \& {Musielak}}]{Cuntz2000ApJ}
{Cuntz}, M., {Saar}, S.~H., \& {Musielak}, Z.~E. 2000, \apjl, 533, L151, \dodoi{10.1086/312609}

\bibitem[{{Curtis} {et~al.}(2019){Curtis}, {Ag{\"u}eros}, {Douglas}, \& {Meibom}}]{Curtis2019ApJ}
{Curtis}, J.~L., {Ag{\"u}eros}, M.~A., {Douglas}, S.~T., \& {Meibom}, S. 2019, \apj, 879, 49, \dodoi{10.3847/1538-4357/ab2393}

\bibitem[{{Curtis} {et~al.}(2020){Curtis}, {Ag{\"u}eros}, {Matt}, {Covey}, {Douglas}, {Angus}, {Saar}, {Cody}, {Vanderburg}, {Law}, {Kraus}, {Latham}, {Baranec}, {Riddle}, {Ziegler}, {Lund}, {Torres}, {Meibom}, {Aguirre}, \& {Wright}}]{Curtis2020ApJ}
{Curtis}, J.~L., {Ag{\"u}eros}, M.~A., {Matt}, S.~P., {et~al.} 2020, \apj, 904, 140, \dodoi{10.3847/1538-4357/abbf58}

\bibitem[{{Cyburt} {et~al.}(2010){Cyburt}, {Amthor}, {Ferguson}, {Meisel}, {Smith}, {Warren}, {Heger}, {Hoffman}, {Rauscher}, {Sakharuk}, {Schatz}, {Thielemann}, \& {Wiescher}}]{Cyburt2010ApJS}
{Cyburt}, R.~H., {Amthor}, A.~M., {Ferguson}, R., {et~al.} 2010, \apjs, 189, 240, \dodoi{10.1088/0067-0049/189/1/240}

\bibitem[{{Damiani} \& {D{\'\i}az}(2016)}]{Damiani2016AA}
{Damiani}, C., \& {D{\'\i}az}, R.~F. 2016, \aap, 589, A55, \dodoi{10.1051/0004-6361/201527100}

\bibitem[{{Damiani} \& {Lanza}(2015)}]{Damiani2015AA}
{Damiani}, C., \& {Lanza}, A.~F. 2015, \aap, 574, A39, \dodoi{10.1051/0004-6361/201424318}

\bibitem[{{Darwin}(1880)}]{Darwin1880RSPT}
{Darwin}, G.~H. 1880, Philosophical Transactions of the Royal Society of London Series I, 171, 713

\bibitem[{{Davenport}(2017)}]{Davenport2017ApJ}
{Davenport}, J. R.~A. 2017, \apj, 835, 16, \dodoi{10.3847/1538-4357/835/1/16}

\bibitem[{{David} {et~al.}(2022){David}, {Angus}, {Curtis}, {van Saders}, {Colman}, {Contardo}, {Lu}, \& {Zinn}}]{David2022ApJ}
{David}, T.~J., {Angus}, R., {Curtis}, J.~L., {et~al.} 2022, \apj, 933, 114, \dodoi{10.3847/1538-4357/ac6dd3}

\bibitem[{{Davoudi} {et~al.}(2021){Davoudi}, {Ba{\c{s}}t{\"u}rk}, {Yal{\c{c}}{\i}nkaya}, {Esmer}, \& {Safari}}]{Davoudi2021AJ}
{Davoudi}, F., {Ba{\c{s}}t{\"u}rk}, {\"O}., {Yal{\c{c}}{\i}nkaya}, S., {Esmer}, E.~M., \& {Safari}, H. 2021, \aj, 162, 210, \dodoi{10.3847/1538-3881/ac1baf}

\bibitem[{{Dawson} \& {Johnson}(2018)}]{DawsonJohnson2018ARAA}
{Dawson}, R.~I., \& {Johnson}, J.~A. 2018, \araa, 56, 175, \dodoi{10.1146/annurev-astro-081817-051853}

\bibitem[{{Dobbs-Dixon} {et~al.}(2004){Dobbs-Dixon}, {Lin}, \& {Mardling}}]{Dobbs-Dixon2004ApJ}
{Dobbs-Dixon}, I., {Lin}, D.~N.~C., \& {Mardling}, R.~A. 2004, \apj, 610, 464, \dodoi{10.1086/421510}

\bibitem[{{Dr{\k{a}}{\.z}kowska} {et~al.}(2023){Dr{\k{a}}{\.z}kowska}, {Bitsch}, {Lambrechts}, {Mulders}, {Harsono}, {Vazan}, {Liu}, {Ormel}, {Kretke}, \& {Morbidelli}}]{Drazkowska2023ASPC}
{Dr{\k{a}}{\.z}kowska}, J., {Bitsch}, B., {Lambrechts}, M., {et~al.} 2023, in Astronomical Society of the Pacific Conference Series, Vol. 534, Protostars and Planets VII, ed. S.~{Inutsuka}, Y.~{Aikawa}, T.~{Muto}, K.~{Tomida}, \& M.~{Tamura}, 717, \dodoi{10.48550/arXiv.2203.09759}

\bibitem[{{Dungee} {et~al.}(2022){Dungee}, {van Saders}, {Gaidos}, {Chun}, {Garc{\'\i}a}, {Magnier}, {Mathur}, \& {Santos}}]{Dungee2022ApJ}
{Dungee}, R., {van Saders}, J., {Gaidos}, E., {et~al.} 2022, \apj, 938, 118, \dodoi{10.3847/1538-4357/ac90be}

\bibitem[{{Efroimsky} \& {Makarov}(2022)}]{Efroimsky2022Univ}
{Efroimsky}, M., \& {Makarov}, V.~V. 2022, Universe, 8, 211, \dodoi{10.3390/universe8040211}

\bibitem[{{Eggleton} {et~al.}(1998){Eggleton}, {Kiseleva}, \& {Hut}}]{Eggleton1998ApJ}
{Eggleton}, P.~P., {Kiseleva}, L.~G., \& {Hut}, P. 1998, \apj, 499, 853, \dodoi{10.1086/305670}

\bibitem[{{El-Badry} {et~al.}(2022){El-Badry}, {Conroy}, {Fuller}, {Kiman}, {van Roestel}, {Rodriguez}, \& {Burdge}}]{El-Badry2022MNRAS}
{El-Badry}, K., {Conroy}, C., {Fuller}, J., {et~al.} 2022, \mnras, 517, 4916, \dodoi{10.1093/mnras/stac2945}

\bibitem[{{Esposito} {et~al.}(2017){Esposito}, {Covino}, {Desidera}, {Mancini}, {Nascimbeni}, {Zanmar Sanchez}, {Biazzo}, {Lanza}, {Leto}, {Southworth}, {Bonomo}, {Su{\'a}rez Mascare{\~n}o}, {Boccato}, {Cosentino}, {Claudi}, {Gratton}, {Maggio}, {Micela}, {Molinari}, {Pagano}, {Piotto}, {Poretti}, {Smareglia}, {Sozzetti}, {Affer}, {Anderson}, {Andreuzzi}, {Benatti}, {Bignamini}, {Borsa}, {Borsato}, {Ciceri}, {Damasso}, {di Fabrizio}, {Giacobbe}, {Granata}, {Harutyunyan}, {Henning}, {Malavolta}, {Maldonado}, {Martinez Fiorenzano}, {Masiero}, {Molaro}, {Molinaro}, {Pedani}, {Rainer}, {Scandariato}, \& {Turner}}]{Espinto2017AA}
{Esposito}, M., {Covino}, E., {Desidera}, S., {et~al.} 2017, \aap, 601, A53, \dodoi{10.1051/0004-6361/201629720}

\bibitem[{{Essick} \& {Weinberg}(2016)}]{Essick2016ApJ}
{Essick}, R., \& {Weinberg}, N.~N. 2016, \apj, 816, 18, \dodoi{10.3847/0004-637X/816/1/18}

\bibitem[{{Ferraz-Mello} {et~al.}(2008){Ferraz-Mello}, {Rodr{\'\i}guez}, \& {Hussmann}}]{Ferraz-Mello2008CeMDA}
{Ferraz-Mello}, S., {Rodr{\'\i}guez}, A., \& {Hussmann}, H. 2008, Celestial Mechanics and Dynamical Astronomy, 101, 171, \dodoi{10.1007/s10569-008-9133-x}

\bibitem[{{Fleming} {et~al.}(2019){Fleming}, {Barnes}, {Davenport}, \& {Luger}}]{Fleming2019ApJ}
{Fleming}, D.~P., {Barnes}, R., {Davenport}, J. R.~A., \& {Luger}, R. 2019, \apj, 881, 88, \dodoi{10.3847/1538-4357/ab2ed2}

\bibitem[{{Fortney} {et~al.}(2021){Fortney}, {Dawson}, \& {Komacek}}]{Fortney2021JGR}
{Fortney}, J.~J., {Dawson}, R.~I., \& {Komacek}, T.~D. 2021, Journal of Geophysical Research (Planets), 126, e06629, \dodoi{10.1029/2020JE006629}

\bibitem[{{Fritzewski} {et~al.}(2021){Fritzewski}, {Barnes}, {James}, \& {Strassmeier}}]{Fritzewski2021AA}
{Fritzewski}, D.~J., {Barnes}, S.~A., {James}, D.~J., \& {Strassmeier}, K.~G. 2021, \aap, 652, A60, \dodoi{10.1051/0004-6361/202140894}

\bibitem[{{Fritzewski} {et~al.}(2023){Fritzewski}, {Barnes}, {Weingrill}, {Granzer}, {Cole-Kodikara}, \& {Strassmeier}}]{Fritzewski2023AA}
{Fritzewski}, D.~J., {Barnes}, S.~A., {Weingrill}, J., {et~al.} 2023, \aap, 674, A152, \dodoi{10.1051/0004-6361/202346083}

\bibitem[{{Fuller}(2017)}]{Fuller2017MNRAS}
{Fuller}, J. 2017, \mnras, 472, 1538, \dodoi{10.1093/mnras/stx2135}

\bibitem[{{Gaia Collaboration} {et~al.}(2018){Gaia Collaboration}, {Brown}, {Vallenari}, {Prusti}, {de Bruijne}, {Babusiaux}, {Bailer-Jones}, {Biermann}, {Evans}, {Eyer}, {Jansen}, {Jordi}, {Klioner}, {Lammers}, {Lindegren}, {Luri}, {Mignard}, {Panem}, {Pourbaix}, {Randich}, {Sartoretti}, {Siddiqui}, {Soubiran}, {van Leeuwen}, {Walton}, {Arenou}, {Bastian}, {Cropper}, {Drimmel}, {Katz}, {Lattanzi}, {Bakker}, {Cacciari}, {Casta{\~n}eda}, {Chaoul}, {Cheek}, {De Angeli}, {Fabricius}, {Guerra}, {Holl}, {Masana}, {Messineo}, {Mowlavi}, {Nienartowicz}, {Panuzzo}, {Portell}, {Riello}, {Seabroke}, {Tanga}, {Th{\'e}venin}, {Gracia-Abril}, {Comoretto}, {Garcia-Reinaldos}, {Teyssier}, {Altmann}, {Andrae}, {Audard}, {Bellas-Velidis}, {Benson}, {Berthier}, {Blomme}, {Burgess}, {Busso}, {Carry}, {Cellino}, {Clementini}, {Clotet}, {Creevey}, {Davidson}, {De Ridder}, {Delchambre}, {Dell'Oro}, {Ducourant}, {Fern{\'a}ndez-Hern{\'a}ndez}, {Fouesneau}, {Fr{\'e}mat}, {Galluccio}, {Garc{\'\i}a-Torres},
  {Gonz{\'a}lez-N{\'u}{\~n}ez}, {Gonz{\'a}lez-Vidal}, {Gosset}, {Guy}, {Halbwachs}, {Hambly}, {Harrison}, {Hern{\'a}ndez}, {Hestroffer}, {Hodgkin}, {Hutton}, {Jasniewicz}, {Jean-Antoine-Piccolo}, {Jordan}, {Korn}, {Krone-Martins}, {Lanzafame}, {Lebzelter}, {L{\"o}ffler}, {Manteiga}, {Marrese}, {Mart{\'\i}n-Fleitas}, {Moitinho}, {Mora}, {Muinonen}, {Osinde}, {Pancino}, {Pauwels}, {Petit}, {Recio-Blanco}, {Richards}, {Rimoldini}, {Robin}, {Sarro}, {Siopis}, {Smith}, {Sozzetti}, {S{\"u}veges}, {Torra}, {van Reeven}, {Abbas}, {Abreu Aramburu}, {Accart}, {Aerts}, {Altavilla}, {{\'A}lvarez}, {Alvarez}, {Alves}, {Anderson}, {Andrei}, {Anglada Varela}, {Antiche}, {Antoja}, {Arcay}, {Astraatmadja}, {Bach}, {Baker}, {Balaguer-N{\'u}{\~n}ez}, {Balm}, {Barache}, {Barata}, {Barbato}, {Barblan}, {Barklem}, {Barrado}, {Barros}, {Barstow}, {Bartholom{\'e} Mu{\~n}oz}, {Bassilana}, {Becciani}, {Bellazzini}, {Berihuete}, {Bertone}, {Bianchi}, {Bienaym{\'e}}, {Blanco-Cuaresma}, {Boch}, {Boeche}, {Bombrun}, {Borrachero},
  {Bossini}, {Bouquillon}, {Bourda}, {Bragaglia}, {Bramante}, {Breddels}, {Bressan}, {Brouillet}, {Br{\"u}semeister}, {Brugaletta}, {Bucciarelli}, {Burlacu}, {Busonero}, {Butkevich}, {Buzzi}, {Caffau}, {Cancelliere}, {Cannizzaro}, {Cantat-Gaudin}, {Carballo}, {Carlucci}, {Carrasco}, {Casamiquela}, {Castellani}, {Castro-Ginard}, {Charlot}, {Chemin}, {Chiavassa}, {Cocozza}, {Costigan}, {Cowell}, {Crifo}, {Crosta}, {Crowley}, {Cuypers}, {Dafonte}, {Damerdji}, {Dapergolas}, {David}, {David}, {de Laverny}, {De Luise}, {De March}, {de Martino}, {de Souza}, {de Torres}, {Debosscher}, {del Pozo}, {Delbo}, {Delgado}, {Delgado}, {Di Matteo}, {Diakite}, {Diener}, {Distefano}, {Dolding}, {Drazinos}, {Dur{\'a}n}, {Edvardsson}, {Enke}, {Eriksson}, {Esquej}, {Eynard Bontemps}, {Fabre}, {Fabrizio}, {Faigler}, {Falc{\~a}o}, {Farr{\`a}s Casas}, {Federici}, {Fedorets}, {Fernique}, {Figueras}, {Filippi}, {Findeisen}, {Fonti}, {Fraile}, {Fraser}, {Fr{\'e}zouls}, {Gai}, {Galleti}, {Garabato}, {Garc{\'\i}a-Sedano}, {Garofalo},
  {Garralda}, {Gavel}, {Gavras}, {Gerssen}, {Geyer}, {Giacobbe}, {Gilmore}, {Girona}, {Giuffrida}, {Glass}, {Gomes}, {Granvik}, {Gueguen}, {Guerrier}, {Guiraud}, {Guti{\'e}rrez-S{\'a}nchez}, {Haigron}, {Hatzidimitriou}, {Hauser}, {Haywood}, {Heiter}, {Helmi}, {Heu}, {Hilger}, {Hobbs}, {Hofmann}, {Holland}, {Huckle}, {Hypki}, {Icardi}, {Jan{\ss}en}, {Jevardat de Fombelle}, {Jonker}, {Juh{\'a}sz}, {Julbe}, {Karampelas}, {Kewley}, {Klar}, {Kochoska}, {Kohley}, {Kolenberg}, {Kontizas}, {Kontizas}, {Koposov}, {Kordopatis}, {Kostrzewa-Rutkowska}, {Koubsky}, {Lambert}, {Lanza}, {Lasne}, {Lavigne}, {Le Fustec}, {Le Poncin-Lafitte}, {Lebreton}, {Leccia}, {Leclerc}, {Lecoeur-Taibi}, {Lenhardt}, {Leroux}, {Liao}, {Licata}, {Lindstr{\o}m}, {Lister}, {Livanou}, {Lobel}, {L{\'o}pez}, {Managau}, {Mann}, {Mantelet}, {Marchal}, {Marchant}, {Marconi}, {Marinoni}, {Marschalk{\'o}}, {Marshall}, {Martino}, {Marton}, {Mary}, {Massari}, {Matijevi{\v{c}}}, {Mazeh}, {McMillan}, {Messina}, {Michalik}, {Millar}, {Molina}, {Molinaro},
  {Moln{\'a}r}, {Montegriffo}, {Mor}, {Morbidelli}, {Morel}, {Morris}, {Mulone}, {Muraveva}, {Musella}, {Nelemans}, {Nicastro}, {Noval}, {O'Mullane}, {Ord{\'e}novic}, {Ord{\'o}{\~n}ez-Blanco}, {Osborne}, {Pagani}, {Pagano}, {Pailler}, {Palacin}, {Palaversa}, {Panahi}, {Pawlak}, {Piersimoni}, {Pineau}, {Plachy}, {Plum}, {Poggio}, {Poujoulet}, {Pr{\v{s}}a}, {Pulone}, {Racero}, {Ragaini}, {Rambaux}, {Ramos-Lerate}, {Regibo}, {Reyl{\'e}}, {Riclet}, {Ripepi}, {Riva}, {Rivard}, {Rixon}, {Roegiers}, {Roelens}, {Romero-G{\'o}mez}, {Rowell}, {Royer}, {Ruiz-Dern}, {Sadowski}, {Sagrist{\`a} Sell{\'e}s}, {Sahlmann}, {Salgado}, {Salguero}, {Sanna}, {Santana-Ros}, {Sarasso}, {Savietto}, {Schultheis}, {Sciacca}, {Segol}, {Segovia}, {S{\'e}gransan}, {Shih}, {Siltala}, {Silva}, {Smart}, {Smith}, {Solano}, {Solitro}, {Sordo}, {Soria Nieto}, {Souchay}, {Spagna}, {Spoto}, {Stampa}, {Steele}, {Steidelm{\"u}ller}, {Stephenson}, {Stoev}, {Suess}, {Surdej}, {Szabados}, {Szegedi-Elek}, {Tapiador}, {Taris}, {Tauran}, {Taylor},
  {Teixeira}, {Terrett}, {Teyssandier}, {Thuillot}, {Titarenko}, {Torra Clotet}, {Turon}, {Ulla}, {Utrilla}, {Uzzi}, {Vaillant}, {Valentini}, {Valette}, {van Elteren}, {Van Hemelryck}, {van Leeuwen}, {Vaschetto}, {Vecchiato}, {Veljanoski}, {Viala}, {Vicente}, {Vogt}, {von Essen}, {Voss}, {Votruba}, {Voutsinas}, {Walmsley}, {Weiler}, {Wertz}, {Wevers}, {Wyrzykowski}, {Yoldas}, {{\v{Z}}erjal}, {Ziaeepour}, {Zorec}, {Zschocke}, {Zucker}, {Zurbach}, \& {Zwitter}}]{Gaia2018AA}
{Gaia Collaboration}, {Brown}, A.~G.~A., {Vallenari}, A., {et~al.} 2018, \aap, 616, A1, \dodoi{10.1051/0004-6361/201833051}

\bibitem[{{Gaia Collaboration} {et~al.}(2023){Gaia Collaboration}, {Vallenari}, {Brown}, {Prusti}, {de Bruijne}, {Arenou}, {Babusiaux}, {Biermann}, {Creevey}, {Ducourant}, {Evans}, {Eyer}, {Guerra}, {Hutton}, {Jordi}, {Klioner}, {Lammers}, {Lindegren}, {Luri}, {Mignard}, {Panem}, {Pourbaix}, {Randich}, {Sartoretti}, {Soubiran}, {Tanga}, {Walton}, {Bailer-Jones}, {Bastian}, {Drimmel}, {Jansen}, {Katz}, {Lattanzi}, {van Leeuwen}, {Bakker}, {Cacciari}, {Casta{\~n}eda}, {De Angeli}, {Fabricius}, {Fouesneau}, {Fr{\'e}mat}, {Galluccio}, {Guerrier}, {Heiter}, {Masana}, {Messineo}, {Mowlavi}, {Nicolas}, {Nienartowicz}, {Pailler}, {Panuzzo}, {Riclet}, {Roux}, {Seabroke}, {Sordo}, {Th{\'e}venin}, {Gracia-Abril}, {Portell}, {Teyssier}, {Altmann}, {Andrae}, {Audard}, {Bellas-Velidis}, {Benson}, {Berthier}, {Blomme}, {Burgess}, {Busonero}, {Busso}, {C{\'a}novas}, {Carry}, {Cellino}, {Cheek}, {Clementini}, {Damerdji}, {Davidson}, {de Teodoro}, {Nu{\~n}ez Campos}, {Delchambre}, {Dell'Oro}, {Esquej},
  {Fern{\'a}ndez-Hern{\'a}ndez}, {Fraile}, {Garabato}, {Garc{\'\i}a-Lario}, {Gosset}, {Haigron}, {Halbwachs}, {Hambly}, {Harrison}, {Hern{\'a}ndez}, {Hestroffer}, {Hodgkin}, {Holl}, {Jan{\ss}en}, {Jevardat de Fombelle}, {Jordan}, {Krone-Martins}, {Lanzafame}, {L{\"o}ffler}, {Marchal}, {Marrese}, {Moitinho}, {Muinonen}, {Osborne}, {Pancino}, {Pauwels}, {Recio-Blanco}, {Reyl{\'e}}, {Riello}, {Rimoldini}, {Roegiers}, {Rybizki}, {Sarro}, {Siopis}, {Smith}, {Sozzetti}, {Utrilla}, {van Leeuwen}, {Abbas}, {{\'A}brah{\'a}m}, {Abreu Aramburu}, {Aerts}, {Aguado}, {Ajaj}, {Aldea-Montero}, {Altavilla}, {{\'A}lvarez}, {Alves}, {Anders}, {Anderson}, {Anglada Varela}, {Antoja}, {Baines}, {Baker}, {Balaguer-N{\'u}{\~n}ez}, {Balbinot}, {Balog}, {Barache}, {Barbato}, {Barros}, {Barstow}, {Bartolom{\'e}}, {Bassilana}, {Bauchet}, {Becciani}, {Bellazzini}, {Berihuete}, {Bernet}, {Bertone}, {Bianchi}, {Binnenfeld}, {Blanco-Cuaresma}, {Blazere}, {Boch}, {Bombrun}, {Bossini}, {Bouquillon}, {Bragaglia}, {Bramante}, {Breedt},
  {Bressan}, {Brouillet}, {Brugaletta}, {Bucciarelli}, {Burlacu}, {Butkevich}, {Buzzi}, {Caffau}, {Cancelliere}, {Cantat-Gaudin}, {Carballo}, {Carlucci}, {Carnerero}, {Carrasco}, {Casamiquela}, {Castellani}, {Castro-Ginard}, {Chaoul}, {Charlot}, {Chemin}, {Chiaramida}, {Chiavassa}, {Chornay}, {Comoretto}, {Contursi}, {Cooper}, {Cornez}, {Cowell}, {Crifo}, {Cropper}, {Crosta}, {Crowley}, {Dafonte}, {Dapergolas}, {David}, {David}, {de Laverny}, {De Luise}, {De March}, {De Ridder}, {de Souza}, {de Torres}, {del Peloso}, {del Pozo}, {Delbo}, {Delgado}, {Delisle}, {Demouchy}, {Dharmawardena}, {Di Matteo}, {Diakite}, {Diener}, {Distefano}, {Dolding}, {Edvardsson}, {Enke}, {Fabre}, {Fabrizio}, {Faigler}, {Fedorets}, {Fernique}, {Fienga}, {Figueras}, {Fournier}, {Fouron}, {Fragkoudi}, {Gai}, {Garcia-Gutierrez}, {Garcia-Reinaldos}, {Garc{\'\i}a-Torres}, {Garofalo}, {Gavel}, {Gavras}, {Gerlach}, {Geyer}, {Giacobbe}, {Gilmore}, {Girona}, {Giuffrida}, {Gomel}, {Gomez}, {Gonz{\'a}lez-N{\'u}{\~n}ez},
  {Gonz{\'a}lez-Santamar{\'\i}a}, {Gonz{\'a}lez-Vidal}, {Granvik}, {Guillout}, {Guiraud}, {Guti{\'e}rrez-S{\'a}nchez}, {Guy}, {Hatzidimitriou}, {Hauser}, {Haywood}, {Helmer}, {Helmi}, {Sarmiento}, {Hidalgo}, {Hilger}, {H{\l}adczuk}, {Hobbs}, {Holland}, {Huckle}, {Jardine}, {Jasniewicz}, {Jean-Antoine Piccolo}, {Jim{\'e}nez-Arranz}, {Jorissen}, {Juaristi Campillo}, {Julbe}, {Karbevska}, {Kervella}, {Khanna}, {Kontizas}, {Kordopatis}, {Korn}, {K{\'o}sp{\'a}l}, {Kostrzewa-Rutkowska}, {Kruszy{\'n}ska}, {Kun}, {Laizeau}, {Lambert}, {Lanza}, {Lasne}, {Le Campion}, {Lebreton}, {Lebzelter}, {Leccia}, {Leclerc}, {Lecoeur-Taibi}, {Liao}, {Licata}, {Lindstr{\o}m}, {Lister}, {Livanou}, {Lobel}, {Lorca}, {Loup}, {Madrero Pardo}, {Magdaleno Romeo}, {Managau}, {Mann}, {Manteiga}, {Marchant}, {Marconi}, {Marcos}, {Marcos Santos}, {Mar{\'\i}n Pina}, {Marinoni}, {Marocco}, {Marshall}, {Martin Polo}, {Mart{\'\i}n-Fleitas}, {Marton}, {Mary}, {Masip}, {Massari}, {Mastrobuono-Battisti}, {Mazeh}, {McMillan}, {Messina}, {Michalik},
  {Millar}, {Mints}, {Molina}, {Molinaro}, {Moln{\'a}r}, {Monari}, {Mongui{\'o}}, {Montegriffo}, {Montero}, {Mor}, {Mora}, {Morbidelli}, {Morel}, {Morris}, {Muraveva}, {Murphy}, {Musella}, {Nagy}, {Noval}, {Oca{\~n}a}, {Ogden}, {Ordenovic}, {Osinde}, {Pagani}, {Pagano}, {Palaversa}, {Palicio}, {Pallas-Quintela}, {Panahi}, {Payne-Wardenaar}, {Pe{\~n}alosa Esteller}, {Penttil{\"a}}, {Pichon}, {Piersimoni}, {Pineau}, {Plachy}, {Plum}, {Poggio}, {Pr{\v{s}}a}, {Pulone}, {Racero}, {Ragaini}, {Rainer}, {Raiteri}, {Rambaux}, {Ramos}, {Ramos-Lerate}, {Re Fiorentin}, {Regibo}, {Richards}, {Rios Diaz}, {Ripepi}, {Riva}, {Rix}, {Rixon}, {Robichon}, {Robin}, {Robin}, {Roelens}, {Rogues}, {Rohrbasser}, {Romero-G{\'o}mez}, {Rowell}, {Royer}, {Ruz Mieres}, {Rybicki}, {Sadowski}, {S{\'a}ez N{\'u}{\~n}ez}, {Sagrist{\`a} Sell{\'e}s}, {Sahlmann}, {Salguero}, {Samaras}, {Sanchez Gimenez}, {Sanna}, {Santove{\~n}a}, {Sarasso}, {Schultheis}, {Sciacca}, {Segol}, {Segovia}, {S{\'e}gransan}, {Semeux}, {Shahaf}, {Siddiqui}, {Siebert},
  {Siltala}, {Silvelo}, {Slezak}, {Slezak}, {Smart}, {Snaith}, {Solano}, {Solitro}, {Souami}, {Souchay}, {Spagna}, {Spina}, {Spoto}, {Steele}, {Steidelm{\"u}ller}, {Stephenson}, {S{\"u}veges}, {Surdej}, {Szabados}, {Szegedi-Elek}, {Taris}, {Taylor}, {Teixeira}, {Tolomei}, {Tonello}, {Torra}, {Torra}, {Torralba Elipe}, {Trabucchi}, {Tsounis}, {Turon}, {Ulla}, {Unger}, {Vaillant}, {van Dillen}, {van Reeven}, {Vanel}, {Vecchiato}, {Viala}, {Vicente}, {Voutsinas}, {Weiler}, {Wevers}, {Wyrzykowski}, {Yoldas}, {Yvard}, {Zhao}, {Zorec}, {Zucker}, \& {Zwitter}}]{Gaia2023AA}
{Gaia Collaboration}, {Vallenari}, A., {Brown}, A.~G.~A., {et~al.} 2023, \aap, 674, A1, \dodoi{10.1051/0004-6361/202243940}

\bibitem[{{Gallet} \& {Bouvier}(2013)}]{Gallet2013AA}
{Gallet}, F., \& {Bouvier}, J. 2013, \aap, 556, A36, \dodoi{10.1051/0004-6361/201321302}

\bibitem[{{Gallet} \& {Bouvier}(2015)}]{Gallet2015AA}
---. 2015, \aap, 577, A98, \dodoi{10.1051/0004-6361/201525660}

\bibitem[{{Gallet} \& {Delorme}(2019)}]{Gallet2019AA}
{Gallet}, F., \& {Delorme}, P. 2019, \aap, 626, A120, \dodoi{10.1051/0004-6361/201834898}

\bibitem[{{Garc{\'\i}a} {et~al.}(2014){Garc{\'\i}a}, {Ceillier}, {Salabert}, {Mathur}, {van Saders}, {Pinsonneault}, {Ballot}, {Beck}, {Bloemen}, {Campante}, {Davies}, {do Nascimento}, {Mathis}, {Metcalfe}, {Nielsen}, {Su{\'a}rez}, {Chaplin}, {Jim{\'e}nez}, \& {Karoff}}]{Garcia2014AA}
{Garc{\'\i}a}, R.~A., {Ceillier}, T., {Salabert}, D., {et~al.} 2014, \aap, 572, A34, \dodoi{10.1051/0004-6361/201423888}

\bibitem[{{Godoy-Rivera} {et~al.}(2021){Godoy-Rivera}, {Pinsonneault}, \& {Rebull}}]{Godoy-Rivera2021ApJS}
{Godoy-Rivera}, D., {Pinsonneault}, M.~H., \& {Rebull}, L.~M. 2021, \apjs, 257, 46, \dodoi{10.3847/1538-4365/ac2058}

\bibitem[{{Goldreich} \& {Lynden-Bell}(1969)}]{GoldreichLyndenbell1969ApJ}
{Goldreich}, P., \& {Lynden-Bell}, D. 1969, \apj, 156, 59, \dodoi{10.1086/149947}

\bibitem[{{Goldreich} \& {Soter}(1966)}]{Goldreich1966Icar}
{Goldreich}, P., \& {Soter}, S. 1966, \icarus, 5, 375, \dodoi{10.1016/0019-1035(66)90051-0}

\bibitem[{{Goodman} \& {Dickson}(1998)}]{Goodman1998ApJ}
{Goodman}, J., \& {Dickson}, E.~S. 1998, \apj, 507, 938, \dodoi{10.1086/306348}

\bibitem[{{Gordon} {et~al.}(2021){Gordon}, {Davenport}, {Angus}, {Foreman-Mackey}, {Agol}, {Covey}, {Ag{\"u}eros}, \& {Kipping}}]{Gordon2021ApJ}
{Gordon}, T.~A., {Davenport}, J. R.~A., {Angus}, R., {et~al.} 2021, \apj, 913, 70, \dodoi{10.3847/1538-4357/abf63e}

\bibitem[{{Gossage} {et~al.}(2023){Gossage}, {Kalogera}, \& {Sun}}]{Gossage2023ApJ}
{Gossage}, S., {Kalogera}, V., \& {Sun}, M. 2023, \apj, 950, 27, \dodoi{10.3847/1538-4357/acc86e}

\bibitem[{{Grevesse} \& {Sauval}(1998)}]{GrevesseSauval1998SSRv}
{Grevesse}, N., \& {Sauval}, A.~J. 1998, \ssr, 85, 161, \dodoi{10.1023/A:1005161325181}

\bibitem[{{Grieves} {et~al.}(2021){Grieves}, {Bouchy}, {Lendl}, {Carmichael}, {Mireles}, {Shporer}, {McLeod}, {Collins}, {Brahm}, {Stassun}, {Gill}, {Bouma}, {Guillot}, {Cointepas}, {Dos Santos}, {Casewell}, {Jenkins}, {Henning}, {Nielsen}, {Psaridi}, {Udry}, {S{\'e}gransan}, {Eastman}, {Zhou}, {Abe}, {Agabi}, {Bakos}, {Charbonneau}, {Collins}, {Colon}, {Crouzet}, {Dransfield}, {Evans}, {Goeke}, {Hart}, {Irwin}, {Jensen}, {Jord{\'a}n}, {Kielkopf}, {Latham}, {Marie-Sainte}, {M{\'e}karnia}, {Nelson}, {Quinn}, {Radford}, {Rodriguez}, {Rowden}, {Schmider}, {Schwarz}, {Smith}, {Stockdale}, {Suarez}, {Tan}, {Triaud}, {Waalkes}, \& {Wingham}}]{Grieves2021AA}
{Grieves}, N., {Bouchy}, F., {Lendl}, M., {et~al.} 2021, \aap, 652, A127, \dodoi{10.1051/0004-6361/202141145}

\bibitem[{{Gruner} \& {Barnes}(2020)}]{GrunerBarnes2020AA}
{Gruner}, D., \& {Barnes}, S.~A. 2020, \aap, 644, A16, \dodoi{10.1051/0004-6361/202038984}

\bibitem[{{Gruner} {et~al.}(2023){Gruner}, {Barnes}, \& {Weingrill}}]{Gruner2023AA}
{Gruner}, D., {Barnes}, S.~A., \& {Weingrill}, J. 2023, \aap, 672, A159, \dodoi{10.1051/0004-6361/202345942}

\bibitem[{{Guillot} {et~al.}(1996){Guillot}, {Burrows}, {Hubbard}, {Lunine}, \& {Saumon}}]{Guillot1996ApJ}
{Guillot}, T., {Burrows}, A., {Hubbard}, W.~B., {Lunine}, J.~I., \& {Saumon}, D. 1996, \apjl, 459, L35, \dodoi{10.1086/309935}

\bibitem[{{Hall} {et~al.}(2021){Hall}, {Davies}, {van Saders}, {Nielsen}, {Lund}, {Chaplin}, {Garc{\'\i}a}, {Amard}, {Breimann}, {Khan}, {See}, \& {Tayar}}]{Hall2021NatAs}
{Hall}, O.~J., {Davies}, G.~R., {van Saders}, J., {et~al.} 2021, Nature Astronomy, 5, 707, \dodoi{10.1038/s41550-021-01335-x}

\bibitem[{{Harre} {et~al.}(2023){Harre}, {Smith}, {Barros}, {Bou{\'e}}, {Csizmadia}, {Ehrenreich}, {Flor{\'e}n}, {Fortier}, {Maxted}, {Hooton}, {Akinsanmi}, {Serrano}, {Ros{\'a}rio}, {Demory}, {Jones}, {Laskar}, {Adibekyan}, {Alibert}, {Alonso}, {Anderson}, {Anglada}, {Asquier}, {B{\'a}rczy}, {Barrado y Navascues}, {Baumjohann}, {Beck}, {Beck}, {Benz}, {Billot}, {Biondi}, {Bonfanti}, {Bonfils}, {Brandeker}, {Broeg}, {Cabrera}, {Cessa}, {Charnoz}, {Collier Cameron}, {Davies}, {Deleuil}, {Delrez}, {Demangeon}, {Erikson}, {Fossati}, {Fridlund}, {Gandolfi}, {Gillon}, {G{\"u}del}, {Hellier}, {Heng}, {Hoyer}, {Isaak}, {Kiss}, {Lecavelier des Etangs}, {Lendl}, {Lovis}, {Luntzer}, {Magrin}, {Nascimbeni}, {Olofsson}, {Ottensamer}, {Pagano}, {Pall{\'e}}, {Persson}, {Peter}, {Piotto}, {Pollacco}, {Queloz}, {Ragazzoni}, {Rando}, {Rauer}, {Ribas}, {Ricker}, {Salmon}, {Santos}, {Scandariato}, {Seager}, {S{\'e}gransan}, {Simon}, {Sousa}, {Steller}, {Szab{\'o}}, {Thomas}, {Udry}, {Ulmer}, {Van Grootel}, {Walton}, {Wilson},
  {Winn}, \& {Wohler}}]{Harre2023AA}
{Harre}, J.~V., {Smith}, A.~M.~S., {Barros}, S.~C.~C., {et~al.} 2023, \aap, 669, A124, \dodoi{10.1051/0004-6361/202244529}

\bibitem[{{Hartman} {et~al.}(2009){Hartman}, {Gaudi}, {Pinsonneault}, {Stanek}, {Holman}, {McLeod}, {Meibom}, {Barranco}, \& {Kalirai}}]{Hartman2009ApJ}
{Hartman}, J.~D., {Gaudi}, B.~S., {Pinsonneault}, M.~H., {et~al.} 2009, \apj, 691, 342, \dodoi{10.1088/0004-637X/691/1/342}

\bibitem[{{Hsieh} \& {Sibuya}(1999)}]{Hsieh1999}
{Hsieh}, P.-F., \& {Sibuya}, Y. 1999, Basic theory of ordinary differential equations (Springer Science \& Business Media)

\bibitem[{{Hut}(1981)}]{Hut1981AA}
{Hut}, P. 1981, \aap, 99, 126

\bibitem[{{Irwin} {et~al.}(2009){Irwin}, {Aigrain}, {Bouvier}, {Hebb}, {Hodgkin}, {Irwin}, \& {Moraux}}]{Irwin2009MNRAS}
{Irwin}, J., {Aigrain}, S., {Bouvier}, J., {et~al.} 2009, \mnras, 392, 1456, \dodoi{10.1111/j.1365-2966.2008.14158.x}

\bibitem[{{Irwin} {et~al.}(2007){Irwin}, {Hodgkin}, {Aigrain}, {Hebb}, {Bouvier}, {Clarke}, {Moraux}, \& {Bramich}}]{Irwin2007MNRAS}
{Irwin}, J., {Hodgkin}, S., {Aigrain}, S., {et~al.} 2007, \mnras, 377, 741, \dodoi{10.1111/j.1365-2966.2007.11640.x}

\bibitem[{{Ivshina} \& {Winn}(2022)}]{Ivshina2022ApJS}
{Ivshina}, E.~S., \& {Winn}, J.~N. 2022, \apjs, 259, 62, \dodoi{10.3847/1538-4365/ac545b}

\bibitem[{{Jackson} {et~al.}(2008){Jackson}, {Greenberg}, \& {Barnes}}]{Jackson2008ApJ}
{Jackson}, B., {Greenberg}, R., \& {Barnes}, R. 2008, \apj, 678, 1396, \dodoi{10.1086/529187}

\bibitem[{{Jontof-Hutter}(2019)}]{Jontof-Hutter2019AREPS}
{Jontof-Hutter}, D. 2019, Annual Review of Earth and Planetary Sciences, 47, 141, \dodoi{10.1146/annurev-earth-053018-060352}

\bibitem[{{Kawaler}(1988)}]{Kawaler1988ApJ}
{Kawaler}, S.~D. 1988, \apj, 333, 236, \dodoi{10.1086/166740}

\bibitem[{{Kunitomo} {et~al.}(2020){Kunitomo}, {Suzuki}, \& {Inutsuka}}]{Kunitomo2020MNRAS}
{Kunitomo}, M., {Suzuki}, T.~K., \& {Inutsuka}, S.-i. 2020, \mnras, 492, 3849, \dodoi{10.1093/mnras/staa087}

\bibitem[{{Kurokawa} \& {Nakamoto}(2014)}]{Kurokawa2014ApJ}
{Kurokawa}, H., \& {Nakamoto}, T. 2014, \apj, 783, 54, \dodoi{10.1088/0004-637X/783/1/54}

\bibitem[{{Lu} {et~al.}(2022){Lu}, {Curtis}, {Angus}, {David}, \& {Hattori}}]{Lu2022aAJ}
{Lu}, Y.~L., {Curtis}, J.~L., {Angus}, R., {David}, T.~J., \& {Hattori}, S. 2022, \aj, 164, 251, \dodoi{10.3847/1538-3881/ac9bee}

\bibitem[{{Lurie} {et~al.}(2017){Lurie}, {Vyhmeister}, {Hawley}, {Adilia}, {Chen}, {Davenport}, {Juri{\'c}}, {Puig-Holzman}, \& {Weisenburger}}]{Lurie2017AJ}
{Lurie}, J.~C., {Vyhmeister}, K., {Hawley}, S.~L., {et~al.} 2017, \aj, 154, 250, \dodoi{10.3847/1538-3881/aa974d}

\bibitem[{{Ma} \& {Fuller}(2021)}]{MaFuller2021ApJ}
{Ma}, L., \& {Fuller}, J. 2021, \apj, 918, 16, \dodoi{10.3847/1538-4357/ac088e}

\bibitem[{{Maciejewski} {et~al.}(2020){Maciejewski}, {Knutson}, {Howard}, {Isaacson}, {Fern{\'a}ndez-Laj{\'u}s}, {DiSisto}, \& {Migaszewski}}]{Maciejewski2020AcA}
{Maciejewski}, G., {Knutson}, H.~A., {Howard}, A.~W., {et~al.} 2020, \actaa, 70, 1, \dodoi{10.32023/0001-5237/70.1.1}

\bibitem[{{Maciejewski} {et~al.}(2018){Maciejewski}, {Fern{\'a}ndez}, {Aceituno}, {Mart{\'\i}n-Ruiz}, {Ohlert}, {Dimitrov}, {Szyszka}, {von Essen}, {Mugrauer}, {Bischoff}, {Michel}, {Mallonn}, {Stangret}, \& {Mo{\'z}dzierski}}]{Maciejewski2018AcA}
{Maciejewski}, G., {Fern{\'a}ndez}, M., {Aceituno}, F., {et~al.} 2018, \actaa, 68, 371, \dodoi{10.32023/0001-5237/68.4.4}

\bibitem[{{Maciejewski} {et~al.}(2022){Maciejewski}, {Fern{\'a}ndez}, {Sota}, {Amado}, {Dimitrov}, {Nikolov}, {Ohlert}, {Mugrauer}, {Bischoff}, {Heyne}, {Hildebrandt}, {Stenglein}, {Ar{\'e}valo}, {Neira}, {Riesco}, {S{\'a}nchez Mart{\'\i}nez}, \& {Verdugo}}]{Maciejewski2022AA}
{Maciejewski}, G., {Fern{\'a}ndez}, M., {Sota}, A., {et~al.} 2022, \aap, 667, A127, \dodoi{10.1051/0004-6361/202244280}

\bibitem[{{Mannaday} {et~al.}(2022){Mannaday}, {Thakur}, {Southworth}, {Jiang}, {Sahu}, {Mancini}, {Va{\v{n}}ko}, {Kundra}, {Gajdo{\v{s}}}, {A-thano}, {Sariya}, {Yeh}, {Griv}, {Mkrtichian}, \& {Shlyapnikov}}]{Mannaday2022AJ}
{Mannaday}, V.~K., {Thakur}, P., {Southworth}, J., {et~al.} 2022, \aj, 164, 198, \dodoi{10.3847/1538-3881/ac91c2}

\bibitem[{{Masuda}(2017)}]{Masuda2017AJ}
{Masuda}, K. 2017, \aj, 154, 64, \dodoi{10.3847/1538-3881/aa7aeb}

\bibitem[{{Mathis}(2015)}]{Mathis2015AA}
{Mathis}, S. 2015, \aap, 580, L3, \dodoi{10.1051/0004-6361/201526472}

\bibitem[{{Mathis} {et~al.}(2016){Mathis}, {Auclair-Desrotour}, {Guenel}, {Gallet}, \& {Le Poncin-Lafitte}}]{Mathis2016AA}
{Mathis}, S., {Auclair-Desrotour}, P., {Guenel}, M., {Gallet}, F., \& {Le Poncin-Lafitte}, C. 2016, \aap, 592, A33, \dodoi{10.1051/0004-6361/201527545}

\bibitem[{{Matsumura} {et~al.}(2010){Matsumura}, {Peale}, \& {Rasio}}]{Matsumura2010ApJ}
{Matsumura}, S., {Peale}, S.~J., \& {Rasio}, F.~A. 2010, \apj, 725, 1995, \dodoi{10.1088/0004-637X/725/2/1995}

\bibitem[{{Matsumura} {et~al.}(2008){Matsumura}, {Takeda}, \& {Rasio}}]{Matsumura2008ApJ}
{Matsumura}, S., {Takeda}, G., \& {Rasio}, F.~A. 2008, \apjl, 686, L29, \dodoi{10.1086/592818}

\bibitem[{{Matt} \& {Pudritz}(2008)}]{MattPudritz2008ApJ}
{Matt}, S., \& {Pudritz}, R.~E. 2008, \apj, 681, 391, \dodoi{10.1086/587453}

\bibitem[{{Matt} {et~al.}(2015){Matt}, {Brun}, {Baraffe}, {Bouvier}, \& {Chabrier}}]{Matt2015ApJ}
{Matt}, S.~P., {Brun}, A.~S., {Baraffe}, I., {Bouvier}, J., \& {Chabrier}, G. 2015, \apjl, 799, L23, \dodoi{10.1088/2041-8205/799/2/L23}

\bibitem[{{Matt} {et~al.}(2012){Matt}, {MacGregor}, {Pinsonneault}, \& {Greene}}]{Matt2012ApJ}
{Matt}, S.~P., {MacGregor}, K.~B., {Pinsonneault}, M.~H., \& {Greene}, T.~P. 2012, \apjl, 754, L26, \dodoi{10.1088/2041-8205/754/2/L26}

\bibitem[{{Mayor} \& {Queloz}(1995)}]{MayorQueloz1995Nat}
{Mayor}, M., \& {Queloz}, D. 1995, \nat, 378, 355, \dodoi{10.1038/378355a0}

\bibitem[{{Mazeh}(2008)}]{Mazeh2008EAS}
{Mazeh}, T. 2008, in EAS Publications Series, Vol.~29, EAS Publications Series, ed. M.~J. {Goupil} \& J.~P. {Zahn}, 1--65, \dodoi{10.1051/eas:0829001}

\bibitem[{{McCormac} {et~al.}(2020){McCormac}, {Gillen}, {Jackman}, {Brown}, {Bayliss}, {Wheatley}, {Anderson}, {Armstrong}, {Bouchy}, {Briegal}, {Burleigh}, {Cabrera}, {Casewell}, {Chaushev}, {Chazelas}, {Chote}, {Cooke}, {Costes}, {Csizmadia}, {Eigm{\"u}ller}, {Erikson}, {Foxell}, {G{\"a}nsicke}, {Goad}, {G{\"u}nther}, {Hodgkin}, {Hooton}, {Jenkins}, {Lambert}, {Lendl}, {Longstaff}, {Louden}, {Moyano}, {Nielsen}, {Pollacco}, {Queloz}, {Rauer}, {Raynard}, {Smith}, {Smalley}, {Soto}, {Turner}, {Udry}, {Vines}, {Walker}, {Watson}, \& {West}}]{McCormac2020MNRAS}
{McCormac}, J., {Gillen}, E., {Jackman}, J. A.~G., {et~al.} 2020, \mnras, 493, 126, \dodoi{10.1093/mnras/staa115}

\bibitem[{{McQuillan} {et~al.}(2013){McQuillan}, {Aigrain}, \& {Mazeh}}]{McQuillan2013aMNRAS}
{McQuillan}, A., {Aigrain}, S., \& {Mazeh}, T. 2013, \mnras, 432, 1203, \dodoi{10.1093/mnras/stt536}

\bibitem[{{McQuillan} {et~al.}(2014){McQuillan}, {Mazeh}, \& {Aigrain}}]{McQuillan2014ApJS}
{McQuillan}, A., {Mazeh}, T., \& {Aigrain}, S. 2014, \apjs, 211, 24, \dodoi{10.1088/0067-0049/211/2/24}

\bibitem[{{Meibom} {et~al.}(2015){Meibom}, {Barnes}, {Platais}, {Gilliland}, {Latham}, \& {Mathieu}}]{Meibom2015Natur}
{Meibom}, S., {Barnes}, S.~A., {Platais}, I., {et~al.} 2015, \nat, 517, 589, \dodoi{10.1038/nature14118}

\bibitem[{{Meibom} {et~al.}(2011){Meibom}, {Barnes}, {Latham}, {Batalha}, {Borucki}, {Koch}, {Basri}, {Walkowicz}, {Janes}, {Jenkins}, {Van Cleve}, {Haas}, {Bryson}, {Dupree}, {Furesz}, {Szentgyorgyi}, {Buchhave}, {Clarke}, {Twicken}, \& {Quintana}}]{Meibom2011ApJ}
{Meibom}, S., {Barnes}, S.~A., {Latham}, D.~W., {et~al.} 2011, \apjl, 733, L9, \dodoi{10.1088/2041-8205/733/1/L9}

\bibitem[{{Messina} {et~al.}(2022){Messina}, {Nardiello}, {Desidera}, {Baratella}, {Benatti}, {Biazzo}, \& {D'Orazi}}]{Messina2022AA}
{Messina}, S., {Nardiello}, D., {Desidera}, S., {et~al.} 2022, \aap, 657, L3, \dodoi{10.1051/0004-6361/202142276}

\bibitem[{{Metcalfe} \& {van Saders}(2017)}]{MetcalfevanSaders2017SoPh}
{Metcalfe}, T.~S., \& {van Saders}, J. 2017, \solphys, 292, 126, \dodoi{10.1007/s11207-017-1157-5}

\bibitem[{{Metcalfe} {et~al.}(2022){Metcalfe}, {Finley}, {Kochukhov}, {See}, {Ayres}, {Stassun}, {van Saders}, {Clark}, {Godoy-Rivera}, {Ilyin}, {Pinsonneault}, {Strassmeier}, \& {Petit}}]{Metcalfe2022ApJ}
{Metcalfe}, T.~S., {Finley}, A.~J., {Kochukhov}, O., {et~al.} 2022, \apjl, 933, L17, \dodoi{10.3847/2041-8213/ac794d}

\bibitem[{{Metcalfe} {et~al.}(2023){Metcalfe}, {Strassmeier}, {Ilyin}, {van Saders}, {Ayres}, {Finley}, {Kochukhov}, {Petit}, {See}, {Stassun}, {Jeffers}, {Marsden}, {Morin}, \& {Vidotto}}]{Metcalfe2023ApJ}
{Metcalfe}, T.~S., {Strassmeier}, K.~G., {Ilyin}, I.~V., {et~al.} 2023, \apjl, 948, L6, \dodoi{10.3847/2041-8213/acce38}

\bibitem[{{Miotello} {et~al.}(2023){Miotello}, {Kamp}, {Birnstiel}, {Cleeves}, \& {Kataoka}}]{Miotello2023ASPC}
{Miotello}, A., {Kamp}, I., {Birnstiel}, T., {Cleeves}, L.~C., \& {Kataoka}, A. 2023, in Astronomical Society of the Pacific Conference Series, Vol. 534, Protostars and Planets VII, ed. S.~{Inutsuka}, Y.~{Aikawa}, T.~{Muto}, K.~{Tomida}, \& M.~{Tamura}, 501, \dodoi{10.48550/arXiv.2203.09818}

\bibitem[{{Morgan} {et~al.}(2024){Morgan}, {Bowler}, {Tran}, {Petigura}, {Nagpal}, \& {Blunt}}]{Morgan2024AJ}
{Morgan}, M., {Bowler}, B.~P., {Tran}, Q.~H., {et~al.} 2024, \aj, 167, 48, \dodoi{10.3847/1538-3881/ad0728}

\bibitem[{{Murray} \& {Dermott}(1999)}]{MurrayDelmott1999}
{Murray}, C.~D., \& {Dermott}, S.~F. 1999, {Solar system dynamics} (Cambridge University Press)

\bibitem[{{Neubauer}(1980)}]{Neubauer1980JGR}
{Neubauer}, F.~M. 1980, \jgr, 85, 1171, \dodoi{10.1029/JA085iA03p01171}

\bibitem[{{Neubauer}(1998)}]{Neubauer1998JGRN}
---. 1998, \jgr, 103, 19843, \dodoi{10.1029/97JE03370}

\bibitem[{{Newton} {et~al.}(2022){Newton}, {Rampalli}, {Kraus}, {Mann}, {Curtis}, {Vanderburg}, {Krolikowski}, {Huber}, {Petter}, {Bieryla}, {Tofflemire}, {Thao}, {Wood}, {Kerr}, {Safanov}, {Strakhov}, {Ciardi}, {Giacalone}, {Dressing}, {Gill}, {Savel}, {Collins}, {Brown}, {Murgas}, {Isogai}, {Narita}, {Palle}, {Quinn}, {Eastman}, {F{\H{u}}r{\'e}sz}, {Shiao}, {Daylan}, {Caldwell}, {Ricker}, {Vanderspek}, {Seager}, {Winn}, {Jenkins}, \& {Latham}}]{Newton2022AJ}
{Newton}, E.~R., {Rampalli}, R., {Kraus}, A.~L., {et~al.} 2022, \aj, 164, 115, \dodoi{10.3847/1538-3881/ac8154}

\bibitem[{{Noyes} {et~al.}(1984){Noyes}, {Weiss}, \& {Vaughan}}]{Noyes1984ApJ}
{Noyes}, R.~W., {Weiss}, N.~O., \& {Vaughan}, A.~H. 1984, \apj, 287, 769, \dodoi{10.1086/162735}

\bibitem[{{Offner} {et~al.}(2023){Offner}, {Moe}, {Kratter}, {Sadavoy}, {Jensen}, \& {Tobin}}]{Offner2023ASPC}
{Offner}, S.~S.~R., {Moe}, M., {Kratter}, K.~M., {et~al.} 2023, in Astronomical Society of the Pacific Conference Series, Vol. 534, Protostars and Planets VII, ed. S.~{Inutsuka}, Y.~{Aikawa}, T.~{Muto}, K.~{Tomida}, \& M.~{Tamura}, 275, \dodoi{10.48550/arXiv.2203.10066}

\bibitem[{{Ogilvie}(2013)}]{Ogilvie2013MNRAS}
{Ogilvie}, G.~I. 2013, \mnras, 429, 613, \dodoi{10.1093/mnras/sts362}

\bibitem[{{Ogilvie}(2014)}]{Ogilvie2014ARAA}
---. 2014, \araa, 52, 171, \dodoi{10.1146/annurev-astro-081913-035941}

\bibitem[{{Ogilvie} \& {Lin}(2007)}]{Ogilvie2007ApJ}
{Ogilvie}, G.~I., \& {Lin}, D.~N.~C. 2007, \apj, 661, 1180, \dodoi{10.1086/515435}

\bibitem[{{Owen}(2019)}]{Owen2019AREPS}
{Owen}, J.~E. 2019, Annual Review of Earth and Planetary Sciences, 47, 67, \dodoi{10.1146/annurev-earth-053018-060246}

\bibitem[{{Patra} {et~al.}(2020){Patra}, {Winn}, {Holman}, {Gillon}, {Burdanov}, {Jehin}, {Delrez}, {Pozuelos}, {Barkaoui}, {Benkhaldoun}, {Narita}, {Fukui}, {Kusakabe}, {Kawauchi}, {Terada}, {Bouma}, {Weinberg}, \& {Broome}}]{Patra2020AJ}
{Patra}, K.~C., {Winn}, J.~N., {Holman}, M.~J., {et~al.} 2020, \aj, 159, 150, \dodoi{10.3847/1538-3881/ab7374}

\bibitem[{{Paxton} {et~al.}(2011){Paxton}, {Bildsten}, {Dotter}, {Herwig}, {Lesaffre}, \& {Timmes}}]{Paxton2011APJS}
{Paxton}, B., {Bildsten}, L., {Dotter}, A., {et~al.} 2011, \apjs, 192, 3, \dodoi{10.1088/0067-0049/192/1/3}

\bibitem[{{Paxton} {et~al.}(2013){Paxton}, {Cantiello}, {Arras}, {Bildsten}, {Brown}, {Dotter}, {Mankovich}, {Montgomery}, {Stello}, {Timmes}, \& {Townsend}}]{Paxton2013APJS}
{Paxton}, B., {Cantiello}, M., {Arras}, P., {et~al.} 2013, \apjs, 208, 4, \dodoi{10.1088/0067-0049/208/1/4}

\bibitem[{{Paxton} {et~al.}(2015){Paxton}, {Marchant}, {Schwab}, {Bauer}, {Bildsten}, {Cantiello}, {Dessart}, {Farmer}, {Hu}, {Langer}, {Townsend}, {Townsley}, \& {Timmes}}]{Paxton2015APJS}
{Paxton}, B., {Marchant}, P., {Schwab}, J., {et~al.} 2015, \apjs, 220, 15, \dodoi{10.1088/0067-0049/220/1/15}

\bibitem[{{Paxton} {et~al.}(2018){Paxton}, {Schwab}, {Bauer}, {Bildsten}, {Blinnikov}, {Duffell}, {Farmer}, {Goldberg}, {Marchant}, {Sorokina}, {Thoul}, {Townsend}, \& {Timmes}}]{Paxton2018APJS}
{Paxton}, B., {Schwab}, J., {Bauer}, E.~B., {et~al.} 2018, \apjs, 234, 34, \dodoi{10.3847/1538-4365/aaa5a8}

\bibitem[{{Paxton} {et~al.}(2019){Paxton}, {Smolec}, {Schwab}, {Gautschy}, {Bildsten}, {Cantiello}, {Dotter}, {Farmer}, {Goldberg}, {Jermyn}, {Kanbur}, {Marchant}, {Thoul}, {Townsend}, {Wolf}, {Zhang}, \& {Timmes}}]{Paxton2019APJS}
{Paxton}, B., {Smolec}, R., {Schwab}, J., {et~al.} 2019, \apjs, 243, 10, \dodoi{10.3847/1538-4365/ab2241}

\bibitem[{{Penev} {et~al.}(2018){Penev}, {Bouma}, {Winn}, \& {Hartman}}]{Penev2018AJ}
{Penev}, K., {Bouma}, L.~G., {Winn}, J.~N., \& {Hartman}, J.~D. 2018, \aj, 155, 165, \dodoi{10.3847/1538-3881/aaaf71}

\bibitem[{{Penev} {et~al.}(2014){Penev}, {Zhang}, \& {Jackson}}]{Penev2014PASP}
{Penev}, K., {Zhang}, M., \& {Jackson}, B. 2014, \pasp, 126, 553, \dodoi{10.1086/677042}

\bibitem[{{Perryman}(2018)}]{Perryman2018}
{Perryman}, M. 2018, {The Exoplanet Handbook} (Cambridge University Press)

\bibitem[{{Rebull} {et~al.}(2017){Rebull}, {Stauffer}, {Hillenbrand}, {Cody}, {Bouvier}, {Soderblom}, {Pinsonneault}, \& {Hebb}}]{Rebull2017ApJ}
{Rebull}, L.~M., {Stauffer}, J.~R., {Hillenbrand}, L.~A., {et~al.} 2017, \apj, 839, 92, \dodoi{10.3847/1538-4357/aa6aa4}

\bibitem[{{Rebull} {et~al.}(2016){Rebull}, {Stauffer}, {Bouvier}, {Cody}, {Hillenbrand}, {Soderblom}, {Valenti}, {Barrado}, {Bouy}, {Ciardi}, {Pinsonneault}, {Stassun}, {Micela}, {Aigrain}, {Vrba}, {Somers}, {Christiansen}, {Gillen}, \& {Collier Cameron}}]{Rebull2016AJ}
{Rebull}, L.~M., {Stauffer}, J.~R., {Bouvier}, J., {et~al.} 2016, \aj, 152, 113, \dodoi{10.3847/0004-6256/152/5/113}

\bibitem[{{Reinhold} \& {Hekker}(2020)}]{ReinholdHekker2020AA}
{Reinhold}, T., \& {Hekker}, S. 2020, \aap, 635, A43, \dodoi{10.1051/0004-6361/201936887}

\bibitem[{{Remus} {et~al.}(2012){Remus}, {Mathis}, \& {Zahn}}]{Remus2012AA}
{Remus}, F., {Mathis}, S., \& {Zahn}, J.~P. 2012, \aap, 544, A132, \dodoi{10.1051/0004-6361/201118160}

\bibitem[{{Ribas} {et~al.}(2015){Ribas}, {Bouy}, \& {Mer{\'\i}n}}]{Ribas2015AA}
{Ribas}, {\'A}., {Bouy}, H., \& {Mer{\'\i}n}, B. 2015, \aap, 576, A52, \dodoi{10.1051/0004-6361/201424846}

\bibitem[{{Ricker} {et~al.}(2014){Ricker}, {Winn}, {Vanderspek}, {Latham}, {Bakos}, {Bean}, {Berta-Thompson}, {Brown}, {Buchhave}, {Butler}, {Butler}, {Chaplin}, {Charbonneau}, {Christensen-Dalsgaard}, {Clampin}, {Deming}, {Doty}, {De Lee}, {Dressing}, {Dunham}, {Endl}, {Fressin}, {Ge}, {Henning}, {Holman}, {Howard}, {Ida}, {Jenkins}, {Jernigan}, {Johnson}, {Kaltenegger}, {Kawai}, {Kjeldsen}, {Laughlin}, {Levine}, {Lin}, {Lissauer}, {MacQueen}, {Marcy}, {McCullough}, {Morton}, {Narita}, {Paegert}, {Palle}, {Pepe}, {Pepper}, {Quirrenbach}, {Rinehart}, {Sasselov}, {Sato}, {Seager}, {Sozzetti}, {Stassun}, {Sullivan}, {Szentgyorgyi}, {Torres}, {Udry}, \& {Villasenor}}]{Ricker2014SPIE}
{Ricker}, G.~R., {Winn}, J.~N., {Vanderspek}, R., {et~al.} 2014, Society of Photo-Optical Instrumentation Engineers (SPIE) Conference Series, Vol. 9143, {Transiting Exoplanet Survey Satellite (TESS)} (Society of Photo-Optical InstrumentationEngineers (SPIE) Conference Series), 914320

\bibitem[{{Ros{\'a}rio} {et~al.}(2022){Ros{\'a}rio}, {Barros}, {Demangeon}, \& {Santos}}]{Rosario2022AA}
{Ros{\'a}rio}, N.~M., {Barros}, S.~C.~C., {Demangeon}, O.~D.~S., \& {Santos}, N.~C. 2022, \aap, 668, A114, \dodoi{10.1051/0004-6361/202244513}

\bibitem[{{Sakurai}(1985)}]{Sakurai1985AA}
{Sakurai}, T. 1985, \aap, 152, 121

\bibitem[{{Santos} {et~al.}(2021){Santos}, {Breton}, {Mathur}, \& {Garc{\'\i}a}}]{Santos2021ApJS}
{Santos}, A.~R.~G., {Breton}, S.~N., {Mathur}, S., \& {Garc{\'\i}a}, R.~A. 2021, \apjs, 255, 17, \dodoi{10.3847/1538-4365/ac033f}

\bibitem[{{Santos} {et~al.}(2019){Santos}, {Garc{\'\i}a}, {Mathur}, {Bugnet}, {van Saders}, {Metcalfe}, {Simonian}, \& {Pinsonneault}}]{Santos2019ApJS}
{Santos}, A.~R.~G., {Garc{\'\i}a}, R.~A., {Mathur}, S., {et~al.} 2019, \apjs, 244, 21, \dodoi{10.3847/1538-4365/ab3b56}

\bibitem[{{Saunders} {et~al.}(2024){Saunders}, {van Saders}, {Lyttle}, {Metcalfe}, {Li}, {Davies}, {Hall}, {Ball}, {Townsend}, {Creevey}, \& {Dodds}}]{Saunders2024ApJ}
{Saunders}, N., {van Saders}, J.~L., {Lyttle}, A.~J., {et~al.} 2024, \apj, 962, 138, \dodoi{10.3847/1538-4357/ad1516}

\bibitem[{{Schou} {et~al.}(1998){Schou}, {Antia}, {Basu}, {Bogart}, {Bush}, {Chitre}, {Christensen-Dalsgaard}, {Di Mauro}, {Dziembowski}, {Eff-Darwich}, {Gough}, {Haber}, {Hoeksema}, {Howe}, {Korzennik}, {Kosovichev}, {Larsen}, {Pijpers}, {Scherrer}, {Sekii}, {Tarbell}, {Title}, {Thompson}, \& {Toomre}}]{Schou1998ApJ}
{Schou}, J., {Antia}, H.~M., {Basu}, S., {et~al.} 1998, \apj, 505, 390, \dodoi{10.1086/306146}

\bibitem[{{See} {et~al.}(2019){See}, {Matt}, {Folsom}, {Boro Saikia}, {Donati}, {Fares}, {Finley}, {H{\'e}brard}, {Jardine}, {Jeffers}, {Lehmann}, {Marsden}, {Mengel}, {Morin}, {Petit}, {Vidotto}, {Waite}, \& {BCool Collaboration}}]{See2019ApJ}
{See}, V., {Matt}, S.~P., {Folsom}, C.~P., {et~al.} 2019, \apj, 876, 118, \dodoi{10.3847/1538-4357/ab1096}

\bibitem[{{Shoda} {et~al.}(2020){Shoda}, {Suzuki}, {Matt}, {Cranmer}, {Vidotto}, {Strugarek}, {See}, {R{\'e}ville}, {Finley}, \& {Brun}}]{Shoda2020ApJ}
{Shoda}, M., {Suzuki}, T.~K., {Matt}, S.~P., {et~al.} 2020, \apj, 896, 123, \dodoi{10.3847/1538-4357/ab94bf}

\bibitem[{{Skumanich}(1972)}]{Skumanich1972ApJ}
{Skumanich}, A. 1972, \apj, 171, 565, \dodoi{10.1086/151310}

\bibitem[{{Soderblom}(2010)}]{Soderblom2010ARAA}
{Soderblom}, D.~R. 2010, \araa, 48, 581, \dodoi{10.1146/annurev-astro-081309-130806}

\bibitem[{{Strugarek}(2016)}]{Strugarek2016ApJ}
{Strugarek}, A. 2016, \apj, 833, 140, \dodoi{10.3847/1538-4357/833/2/140}

\bibitem[{{Strugarek} {et~al.}(2017){Strugarek}, {Bolmont}, {Mathis}, {Brun}, {R{\'e}ville}, {Gallet}, \& {Charbonnel}}]{Strugarek2017ApJ}
{Strugarek}, A., {Bolmont}, E., {Mathis}, S., {et~al.} 2017, \apjl, 847, L16, \dodoi{10.3847/2041-8213/aa8d70}

\bibitem[{{Strugarek} {et~al.}(2014){Strugarek}, {Brun}, {Matt}, \& {R{\'e}ville}}]{Strugarek2014ApJ}
{Strugarek}, A., {Brun}, A.~S., {Matt}, S.~P., \& {R{\'e}ville}, V. 2014, \apj, 795, 86, \dodoi{10.1088/0004-637X/795/1/86}

\bibitem[{{Suzuki} {et~al.}(2013){Suzuki}, {Imada}, {Kataoka}, {Kato}, {Matsumoto}, {Miyahara}, \& {Tsuneta}}]{Suzuki2013PASJ}
{Suzuki}, T.~K., {Imada}, S., {Kataoka}, R., {et~al.} 2013, \pasj, 65, 98, \dodoi{10.1093/pasj/65.5.98}

\bibitem[{{Tayar} {et~al.}(2022){Tayar}, {Claytor}, {Huber}, \& {van Saders}}]{Tayar2022ApJ}
{Tayar}, J., {Claytor}, Z.~R., {Huber}, D., \& {van Saders}, J. 2022, \apj, 927, 31, \dodoi{10.3847/1538-4357/ac4bbc}

\bibitem[{{Tejada Arevalo} {et~al.}(2021){Tejada Arevalo}, {Winn}, \& {Anderson}}]{Tejada-Arevalo2021ApJ}
{Tejada Arevalo}, R.~A., {Winn}, J.~N., \& {Anderson}, K.~R. 2021, \apj, 919, 138, \dodoi{10.3847/1538-4357/ac1429}

\bibitem[{{Tokuno} {et~al.}(2023){Tokuno}, {Suzuki}, \& {Shoda}}]{Tokuno2023MNRAS}
{Tokuno}, T., {Suzuki}, T.~K., \& {Shoda}, M. 2023, \mnras, 520, 418, \dodoi{10.1093/mnras/stad103}

\bibitem[{{Turner} {et~al.}(2021){Turner}, {Ridden-Harper}, \& {Jayawardhana}}]{Turner2021AJ}
{Turner}, J.~D., {Ridden-Harper}, A., \& {Jayawardhana}, R. 2021, \aj, 161, 72, \dodoi{10.3847/1538-3881/abd178}

\bibitem[{{van Saders} {et~al.}(2016){van Saders}, {Ceillier}, {Metcalfe}, {Silva Aguirre}, {Pinsonneault}, {Garc{\'\i}a}, {Mathur}, \& {Davies}}]{vanSaders2016Nature}
{van Saders}, J.~L., {Ceillier}, T., {Metcalfe}, T.~S., {et~al.} 2016, \nat, 529, 181, \dodoi{10.1038/nature16168}

\bibitem[{{Weber} \& {Davis}(1967)}]{WeberDavis1967ApJ}
{Weber}, E.~J., \& {Davis}, Leverett, J. 1967, \apj, 148, 217, \dodoi{10.1086/149138}

\bibitem[{{Wheatley} {et~al.}(2018){Wheatley}, {West}, {Goad}, {Jenkins}, {Pollacco}, {Queloz}, {Rauer}, {Udry}, {Watson}, {Chazelas}, {Eigm{\"u}ller}, {Lambert}, {Genolet}, {McCormac}, {Walker}, {Armstrong}, {Bayliss}, {Bento}, {Bouchy}, {Burleigh}, {Cabrera}, {Casewell}, {Chaushev}, {Chote}, {Csizmadia}, {Erikson}, {Faedi}, {Foxell}, {G{\"a}nsicke}, {Gillen}, {Grange}, {G{\"u}nther}, {Hodgkin}, {Jackman}, {Jord{\'a}n}, {Louden}, {Metrailler}, {Moyano}, {Nielsen}, {Osborn}, {Poppenhaeger}, {Raddi}, {Raynard}, {Smith}, {Soto}, \& {Titz-Weider}}]{Wheatley2018MNRAS}
{Wheatley}, P.~J., {West}, R.~G., {Goad}, M.~R., {et~al.} 2018, \mnras, 475, 4476, \dodoi{10.1093/mnras/stx2836}

\bibitem[{{Winn} {et~al.}(2010){Winn}, {Fabrycky}, {Albrecht}, \& {Johnson}}]{Winn2010ApJ}
{Winn}, J.~N., {Fabrycky}, D., {Albrecht}, S., \& {Johnson}, J.~A. 2010, \apjl, 718, L145, \dodoi{10.1088/2041-8205/718/2/L145}

\bibitem[{{Winn} \& {Fabrycky}(2015)}]{Winn2015ARAA}
{Winn}, J.~N., \& {Fabrycky}, D.~C. 2015, \araa, 53, 409, \dodoi{10.1146/annurev-astro-082214-122246}

\bibitem[{{Wong} {et~al.}(2022){Wong}, {Shporer}, {Vissapragada}, {Greklek-McKeon}, {Knutson}, {Winn}, \& {Benneke}}]{Wong2022AJ}
{Wong}, I., {Shporer}, A., {Vissapragada}, S., {et~al.} 2022, \aj, 163, 175, \dodoi{10.3847/1538-3881/ac5680}

\bibitem[{{Wright} {et~al.}(2011){Wright}, {Drake}, {Mamajek}, \& {Henry}}]{Wright2011ApJ}
{Wright}, N.~J., {Drake}, J.~J., {Mamajek}, E.~E., \& {Henry}, G.~W. 2011, \apj, 743, 48, \dodoi{10.1088/0004-637X/743/1/48}

\bibitem[{{Xue} {et~al.}(2014){Xue}, {Suto}, {Taruya}, {Hirano}, {Fujii}, \& {Masuda}}]{Xue2014ApJ}
{Xue}, Y., {Suto}, Y., {Taruya}, A., {et~al.} 2014, \apj, 784, 66, \dodoi{10.1088/0004-637X/784/1/66}

\bibitem[{{Yee} {et~al.}(2020){Yee}, {Winn}, {Knutson}, {Patra}, {Vissapragada}, {Zhang}, {Holman}, {Shporer}, \& {Wright}}]{Yee2020ApJL}
{Yee}, S.~W., {Winn}, J.~N., {Knutson}, H.~A., {et~al.} 2020, \apjl, 888, L5, \dodoi{10.3847/2041-8213/ab5c16}

\bibitem[{{Zahn}(1966)}]{Zahn1966AnAp}
{Zahn}, J.~P. 1966, Annales d'Astrophysique, 29, 489

\bibitem[{{Zahn}(1975)}]{Zahn1975AA}
---. 1975, \aap, 41, 329

\bibitem[{{Zahn}(1977)}]{Zahn1977AA}
---. 1977, \aap, 57, 383

\bibitem[{{Zahn} \& {Bouchet}(1989)}]{Zahn1989AA}
{Zahn}, J.~P., \& {Bouchet}, L. 1989, \aap, 223, 112

\end{thebibliography}
\bibliographystyle{aasjournal}



\end{document}